\documentclass[11pt]{IEEEtran11}
\twocolumn

\newcommand{\commentout}[1]{}

\include{epsf}

\newcommand{\calevts}{\theta}
\newcommand{\arevt}{r}

\newcommand\BT{\begin{tabbing}}
\newcommand\ET{\end{tabbing}}

\renewcommand{\Lambda}{S}
\newcommand{\accept}{\mbox{\em accept}}
\newcommand{\reject}{\mbox{\em reject}}

\begin{document}
\title{\Large\bf A Decision-Theoretic Approach to Resource Allocation in
Wireless Multimedia Networks}
\author{ }
\author{Zygmunt Haas, Joseph Y. Halpern, Li Li, Stephen B. Wicker \\
School of Electrical Engineering/Computer Science Dept.\\
Cornell University \\
Ithaca, New York 14853 \\
{\tt haas@ee.cornell.edu, halpern@cs.cornell.edu,} \\
{lili@cs.cornell.edu, wicker@ee.cornell.edu}}

\date{}

\maketitle
\begin{abstract}
The allocation of scarce spectral resources to support as many user
applications as possible while maintaining reasonable quality of service
is a fundamental problem in wireless communication.  We argue that the
problem is best formulated in terms of decision theory.  We propose a
scheme that takes decision-theoretic concerns (like preferences)
into account and discuss the difficulties and subtleties involved in
applying standard techniques from the theory of Markov Decision
Processes (MDPs) in constructing an algorithm that is
decision-theoretically optimal. As an example of the proposed framework,
we construct such an algorithm under some
simplifying assumptions.
Additionally, we present
analysis and simulation results that show that our algorithm meets its
design
goals. Finally, we investigate how far from optimal
one
well-known
heuristic is.
The main contribution of our results is in providing insight
and guidance for the design of near-optimal admission-control policies.

\end{abstract}

\section{Introduction}
\label{sec-introduction}
When a user
makes a call in a wireless network, the system must decide whether to
admit the call or block it.%
\footnote{In this paper, we use the term ``user'' to refer to the actual
user or the application running on the user's mobile device; we use the
generic term ``call'' to indicate an attempt to initiate a session,
which could be of any medium type or types (e.g., voice, data, video,
hybrid); we use the term ``block'' to mean rejecting new calls,
the term ``drop'' to mean
rejecting calls already in progress (e.g., during a handoff process),
and the term ``reject'' to mean either block or drop.}
 What makes the decision difficult is that,
to achieve some sense of
optimality, one needs to consider the future  status of the network
resources and the pattern of the
future arrival requests. For example, even if the network currently has
sufficient resources to handle
the call, admitting the call may result in other, already in-progress calls
being dropped in the future. 
This problem is
aggravated by 
the trend to decreasing cell sizes
coupled with an increased demand for
multimedia services.  
The reduction in cell size leads to increased spectrum-utilization efficiency,
but also increases the number
of handoffs experienced during a typical connection.  
This means that the call requests at one cell 
will be more affected by decisions made in nearby cells.
Furthermore,
as the number of cells increases, the amount of resources per cell
decreases.
Thus, the {\em trunking efficiency\/} is reduced, leading to more severe
fluctuations 
in the quality of service (QoS). The provision of connection-level
QoS raises a distinct yet dependent set of issues. Guarantees made by one cell
place future burdens, due to handoffs, on the resources of other cells.

This particular problem is but one of many in the general area of
admission-control policies.  The
crafting of admission policies is generally approached by focusing on a
subset of the design issues, while ignoring others. For example, several
schemes for wireless networks
\cite{ChoiShinSIG98,NagSchw96,LuBharg96} have been
proposed to provide connection-level QoS
by basing the call-admission decision
on a requirement that the
handoff-dropping probability be kept below a certain level.
Clearly, however, lowering the
handoff-dropping probability will, in
general, mean increasing the probability of calls being blocked.  A
major issue remains unaddressed: is
it worth decreasing the handoff-dropping probability from, say, 2\% to 1\%
if it means increasing the
probability of blocking from 5\% to 10\%? In other words, what is an
acceptable tradeoff between call blocking and call dropping probabilities?

Tolerance to blocking and dropping among users 
is also not addressed by a simple 
minimization of handoff-dropping probability.  
For example, compare a
typical voice call with a file transfer.  A user making a voice call may be
mildly frustrated if she
cannot connect to the server, but would be much more annoyed if the call
were dropped halfway through the conversation.
On the other hand, a user attempting to do 
a number of FTP file transfers may prefer to be
connected immediately, even if this means a higher probability of being
dropped, since this increases the
probability of 
at least some rapid file transfers.
Should the
network treat voice and file
transfer connections uniformly and provide a single dropping probability
for both?

We believe that sensible admission
decisions should take both utilities and probabilities into account, and
are best formulated in decision-theoretic terms.
In this paper, we describe a general decision-theoretic approach to
call admission.
Although we use a specific example to demonstrate our approach, 
our approach gives
a framework that can accommodate various QoS
requirements in the multimedia resource-allocation process.
Our focus is
on wireless networks, but the generalization to other contexts is
immediate.  We show that even a
simplistic decision-theoretic approach to wireless network admission
policies produces substantial performance improvement
over several well-known heuristics.

The key ingredients in decision theory are {\em probability\/} and
{\em utility}. Probability is a measure of the
likelihood of an event, while the utility (which can be interpreted as
the reward) measures the event's
perceived importance.  A decision-theoretic admission-control policy
can take into account
the relative utility of blocking and dropping,
as well as other
considerations, such as network utilization or
service differentiation
and delay requirements.

If we are to use a decision-theoretic approach, we must somehow determine the
probabilities and utilities.  We assume that the probabilities can be
obtained by taking sufficient statistical measurements.  Obtaining the
utilities is somewhat more problematic.
Utilities are subjective and different users may have
different utilities for the same application.  Moreover, there is the
problem of what is known as {\em inter-subjective\/} utility: one user's
utility of 10 may not be equivalent to another user's 10.

We sidestep these problems to some extent by assuming that we have only
one user, which can be thought of as the system provider.
Of course,
it is in the system provider's interest to keep users
as happy as possible. So we can assume that, to some extent, the system
provider's utility
is representative of a typical user's utilities.
This is
particularly true if, as we would expect, the system provider gives
higher utilities to calls for which users are prepared to pay more.

To use our approach, the system provider must 
decide what the relative utilities of blocking vs.~dropping ought to be.
Presumably this will be done by weighing the profits lost from
blocking a call, compared to the profits lost
due to customers switching to a different provider as a result of having
calls dropped too frequently
(and possibly refunds due to dropped calls).
In this paper, we assume that we are
simply given the relevant utilities, while recognizing that obtaining
them may be a nontrivial problem.

Once we have the relevant probabilities and utilities,
we can employ Markov Decision Processes (MDPs) \cite{Puterman94}.
More specifically, we model the call-admission problem as an MDP, allowing us to use
well-known techniques for finding the optimal
call-admission policy.
However, we must be careful in modeling the call-admission
problem as an MDP.  The standard techniques for finding the optimal
policy in an MDP run in time polynomial in the size of the state space
and the number of actions.
A larger state space leads to a more accurate model of the system. But,
if we are not careful, the state
space can quickly become unmanageable.  We discuss this issue in detail
and show that under some
simplifying and yet reasonably practical assumptions
we can construct a manageable state space.

We are not the first to apply MDPs to the call-admission problem (see
Section~\ref{sec-related_work} for other references).  
However, there are a number of subtleties that arise in using MDPs in
this context that do not seem to have been considered earlier.  
For example, 
it is typically assumed that we are given the probabilities and utilities in an
MDP, and that we must then construct an optimal admission
policy.  However, in a wireless network, the probabilities are
themselves influenced by the admission
policy we construct.  For example, the probability that a call will be
handed off from a neighboring cell depends on the probability that the
neighboring cell will accept the call.  This, in turn, depends on the
admission policy, which was
developed using assumptions about those probabilities.
This is an issue that does not seem to have been considered in the
literature on MDPs,
perhaps because it was assumed that the probabilities of call blocking and
dropping were so low that they could be ignored.  Since we want our system
to be able to deal with situations where the probability of blocking may
be 
nonnegligible,
we must address this problem as part of our solution.
We do so by constructing a
sequence of policies that converge to a limit.  In the limit, we have a
policy that is
``locally''
optimal
with respect to a given probability measure and utility function, such
that the probability measure is precisely that induced by the policy.
By finding a (locally) optimal policy using MDPs, we are able to make
optimal tradeoffs between such concerns as providing the desired
connection-level QoS and spectrum utilization.
We show by an extensive series of simulations that our approach
provides substantial performance improvement 
over one of the best heuristic approaches
considered in the literature,
given some standard assumptions about arrival times and holding times
of calls.
However, we do not view our major contribution as the performance
improvements in this particular case.
Rather, we consider our major contribution as the general
decision-theoretic framework
and the discussion regarding how to go about using it in practice.
We believe that this framework provides us
with a general methodology for evaluating admission-control
policies and
for characterizing optimal policies.
The rest of this paper is organized as follows.
Section \ref{sec-network_model_and_QoS_classes}
discusses
our assumptions about the underlying system and
our use of QoS classes.
Section \ref{sec-formulating_admission_control}
considers how the admission-control problem can be formulated as an MDP
and the subtleties involved in doing so.
Section \ref{sec-computing_optimal_policy} shows how the optimal
admission-control policy can be computed.
Section \ref{sec-experiments_with2QoS}
presents 
results from an extensive series of simulations that compares the
performance of the decision-theoretic
admission policy for the case of two QoS classes, one for data and one
for video, with 
a well-known
heuristic approach \cite{ChoiShinMobi98,NagSchw96}.  Related
work is described in
Section \ref{sec-related_work},
which also offers some concluding remarks.

\section{Network Model and QoS Classes}
\label{sec-network_model_and_QoS_classes}
We assume that the network
consists of a cellular wireless portion and a wired
backbone. The wireless portion consists of 
a set of cells,
each cell contains a single base station (BS). All the
BSs are connected to the same Mobile Switching Center (MSC). The MSC is
responsible for switching
calls
between cells and for maintaining
connections to the wired backbone.
We focus solely on the wireless portion in this
paper; an integrated admission-control scheme can be obtained by
taking the wired portion into account.

We want to allow the network the possibility of treating different
calls in different ways
due to, for instance, user preferences and tolerances.
We would certainly expect different traffic classes
(e.g., voice, video, data)
to be treated differently, but, as we observed earlier, we may
also want to treat calls within the same traffic class in
different ways.
Thus, we propose that each traffic class be partitioned into a number
of QoS classes, so that the network provides $K$ QoS classes altogether.
We may
further want to refine each QoS class into a number of layers.
The more layers assigned to a traffic stream, the better QoS is provided.
Partitions into layers
give the system an option for further QoS differentiation beyond just
the decision of whether or not to admit a call.
This would, for instance, allow us to take advantage of recent advances
in video
coding \cite{amv96,McCanne96} by representing video (or audio) streams as
multi-layer scalable flows that
can adapt dynamically to changing network and link conditions.
Dealing with layers is a straightforward extension of the ideas
presented here;
for ease of exposition, we do not consider layers further in this
paper.

We assume that each QoS class is associated with 
a set of numbers
representing utilities: the utility
(cost)
per unit of time of
handling a call in that service class,
the negative utility (cost) of blocking a call,
and the cost of dropping a call during a handoff.
(The idea of associating a set of utilities with
a QoS class was inspired by the QoS contract notion
in \cite{AbdelShin98}.)
\section{Formulating Admission Control as an MDP}
\label{sec-formulating_admission_control}

Our goal is to construct an
admission-control policy that decides, for each new call and
handoff call, whether it should be admitted or rejected and, if it is
admitted,
at what QoS level.  The
possibility of queuing incoming calls for later admission is not
considered in this paper.  We want to find a policy that maximizes
expected utility.

It is not sufficient to simply maximize the
expected utility of current new call requests and current admitted
calls.  This does not take into account the future need for handoffs as
active users move across cell boundaries.
Such an approach would result in future expected utility being much
lower than the current expected utility.
Instead, we try to maximize the expected {\em total utility} (in a
sense made precise below).  This means we must consider
utility streams as opposed to temporally isolated utility values.

We cannot just identify the total utility with the sum of utilities over
time,
since this is in
general an infinite sum.  There are two standard approaches in the
literature to
defining total utility \cite{Puterman94}.
The first is to discount the utility over time by some factor
$\gamma$, with $0 < \gamma < 1$.
That is, if $u_k$ is the utility obtained at time $k$, we take the total
utility to be $\sum_{k=1}^{\infty} \gamma^k u_k$. We then try to find
the policy that maximizes the expected total utility.
This definition is meant to
capture the notion of present value in accounting: a dollar now
is worth more than a dollar one time unit later.
The smaller $\gamma$ is, the more we weight the present.
This is reasonable if the time units are reasonably large.
For example, banks pay interest so that \$95 at year $t$ can become
\$100 at year $t+1$.  This suggests that having \$100 at year $t+1$ is
the same as having \$95 at year $t$; i.e., we can assign $\gamma = .95$ to
capture this effect.

A second approach to compute the total utility seems
more appropriate in our context, since we are dealing with small time
intervals.
The focus in this approach is the average utility per unit of time;
then
the total utility over $M$ time units can just be taken to be $M$ times
the average utility.  Thus, in this approach we try to maximize the
expected value of
$\lim_{M \rightarrow \infty} (\sum_{k=1}^M u_i)/M$.
Whichever approach in defining the total utility we use, we can formulate the problem
of finding the policy which optimizes the expected total utility as a
{\em Markov Decision Process} (MDP).  Formally,
an MDP is a tuple $(S,A,T,R)$ consisting of a set $S$ of states, a set
$A$ of actions,
a {\em transition function\/} $T: S \times A \rightarrow
\Pi(S)$, where $\Pi(S)$ is the set of probability distributions over
$S$, and a {\em reward function\/} $R: S \times A \rightarrow {\bf R}$,
where ${\bf R}$ is the set of real numbers.  
Intuitively, $T(s,a)$
describes the probability of ending up in a state $s' \in S$ if we
perform action $a$ in state $s$, while $R(s,a)$ describes the immediate
reward obtained if we perform action $a$ in state $s$.
There are standard techniques for finding the
optimal policy in MDPs, such as value iteration and policy iteration.
These algorithms and a number of their variants have been studied in
detail, and it is well understood what are the relative advantages of
each algorithm (again, see \cite{Puterman94}).
We want to model the call-admission problem as an MDP, so as to be able
to use all the power of MDPs.
We now discuss how this can be done, considering the components of an
MDP one by one.  We pay particular attention to the problem of keeping
the size of the state space manageable.
 Our discussion here is at a
general level; 
we return to the specific issues raised here when we consider
simulation results in Section~\ref{sec-experiments_with2QoS}.
\subsection{The State Space}
We represent the state of the cell as the number of calls in
progress in the cell.
Decisions at a cell are made on the basis of this state.
Clearly,
the more information about the system we put into the state, the more
accurate the model of the system is, and thus, the more accurate our estimation of the
actual rewards received.
For our algorithms to work, we also need to put enough information into
the state to guarantee the {\em Markov property}; i.e., we want the
probability of making a transition from one state to another to depend
only on the current state, not on the previous history.
On the other hand,
the more details we put into
the state, the more states there are, and the longer it will take to
compute the optimal policy.  If we put in too much details, the state
space will be so large that we cannot compute the optimal policy 
in a reasonable length of time.  Thus, there is a tradeoff that must be
made when modeling the state; we may have to model
only certain features, while ignoring others.
A technique
such as value iteration computes the optimal policy with respect to
the state model.
Of course, if our model of the state does not capture enough features of relevance
to the problem, the optimal policy computed by the MDP 
may not be optimal in practice.

We briefly discuss some of these issues
that arise in modeling the state here and the choices we made in our
simulations.
\begin{itemize}
\item Should we include time in the state?  Clearly, on one hand, traffic and mobility
patterns depend on time of day (and day of the week and week of the
year), so we would get a more accurate model by including time. On the
other hand, adding time as a separate component in the state space
means that the number of states is increased by a factor of $T$, where
$T$ is the total number of time units that we deem relevant.  (For
example, if we decide that all that is relevant is the hour of the day,
then there would be 24 time units; if we want to model the behavior at the
level of minutes, then we would have 1440 time units.)
It seems to us that, in practice, the overhead of including time is not
worth the ensuing computational complexity, and we are better off having a
number of separate MDPs, one for each expected pattern, and solving each
of them separately. (Thus, for example, we could have a
``mid-day'' MDP, a middle-of-the-night MDP, and ``morning rush hour''
MDP, and so on.)

\item Should we include information about neighboring cells in the
state?
The number of calls in neighboring cells (together with the mobility
model) in fact determines the handoff probability
(the distribution that describes the probability that a call will be
handed off in the next $t$ time units).
The handoff probability
is in fact necessary for even defining
the MDP, since it affects the transition probability.
On the other hand, keeping track of this information can
significantly increase the size of the state space.
Fortunately, in many cases of interest, it is possible to estimate the
handoff probability (and thus the transition probability), although
subtleties arise that seem to have been ignored in other papers on the
subject.  We return to this issue in
Section~\ref{sec-computing_optimal_policy}.
\item Should we keep track of how long each call has been in the
system?
There are (at least) two reasons why such information may be necessary.
For one, it may be
necessary
in order to
estimate
how much longer the call will remain in the system.  
(For example, if the length of a call is characterized by a heavy-tailed
distribution, then the probability that a call
terminates depends on how long it has lasted.)
For another, if the utility of a call is
a nonlinear function of the length of the call, it will be necessary to
include this information in the state in order to correctly calculate
the utilities.   In our simulations, neither of these reasons apply.
Our assumptions guarantee that the probability of a call termination is
independent of how long it has been in progress 
(i.e. the Markovian property)
and our utility does not
depend in a nonlinear way on the length of the call.
\item Do we have to include in the state information about the
most recent call
(what QoS class it is in, whether it is a new call or a handoff, etc.)?
We have decided to include this information; the alternative would be to
have a much more complicated choice of actions.  We return to this issue
in Section~\ref{sec:act}.
\end{itemize}
Given these considerations, we take the state space
$S$ to consist of
vectors of length $K+1$.
(Recall that $K$ is the number of QoS classes.)
We set $s = (x_1, \ldots, x_K,\calevts) \in S$, where $x_i$ is
the number of ongoing calls
in QoS class $i$, for $i=1, \ldots, K$, and $\calevts$ is a description of the
{\em call event\/} that happens at the current time unit if there is
one, and $n$ (for no event), otherwise.
There are three types of call events: a new-call arrival of QoS class
$i$, a handoff-call arrival of QoS class $i$, and call departure of QoS
class $i$.
We describe these events by pairs of the form $(\arevt,i)$, $(h,i)$, and
$(d,i)$, respectively.  We assume that at most one call event happens at
each time.   Thus, there are at most $3K+1$ possible values of this last
component.  (The $(d,i)$ event cannot occur if there are no calls of QoS
class $i$ currently in the cell.)  The state space is constrained
by the total number of channels in the system.
We assume that each call in QoS class $i$ requires $b_i$ channels and
that there are $N$ channels altogether.
Note that ``channels'' are not necessarily  physical channels; they
can be logical channels as well (for instance, the number of
concurrent users in CDMA).
Thus,
$\Lambda=\{s = (x_1, \ldots, x_K, \calevts): x_{i}\geq 0, \,
\sum_{i=1}^{K}b_{i} x_{i} \leq N \}.$
\subsection{The Action Space and Transition Function}\label{sec:act}
Given our representation of states, we need only two possible actions:
{\em accept\/} and {\em reject}.  These actions have the
obvious effect if there is a call arrival at the current time unit, and
no effect at all if there is a call departure or no call event at the
current time unit.
This is captured by the transition function.
Note that the effect of an action on the first $k$ components of a state
is completely deterministic; the only uncertainty is what the next call
event will be.  For example, if $K=2$, then in state $(2,3,(\arevt,2))$, {\em
accept\/} results in a state of the form $(2,4,\calevts)$ while reject results
in a state of the form $(2,3,\calevts)$.  Similarly, in a state of the form
$(2,3,(d,2))$, both accept and reject result in a state of the form
$(2,2,\calevts)$.   The relative probability of each of
these outcomes depends on our assumptions about dwell times,
mobility (how often calls leave one cell for another), and the
probability of call arrival.
It may also depend on the policy itself, since the policy affects the
number of calls in each cell, which in turn may affect the handoff
probability.  For example, if the policy rejects all calls, then the
handoff probability is guaranteed to be 0.  We return to this issue in
Sections~\ref{sec-computing_optimal_policy}
and~\ref{sec-experiments_with2QoS}.

We can represent the transition function in terms of two matrices, one
for the action {\em accept\/} and one for {\em reject}.  The $(s,t)$
entry
of the {\em accept\/} matrix describes the probability of going from $s$
to $t$ if the {\em accept\/} action is performed, and similarly for {\em
reject}.  Since there are only $3K+1$ entries in each row that have
positive probability, the matrix is relatively sparse.  This may make
it possible to speed up some of the computations involved.

As we said earlier, we could use a state representation that
did not include the last component.  We would then need actions of the
form $(a_1, \ldots, a_{2K})$, where each $a_i$ is either $\accept$ or
$\reject$. For $i = 1, \ldots, K$, the $a_i$ component tells us what to
do if the next call event is an arrival of a new call of QoS class $i$;
for $i = K + j$, $j = 1, \ldots, K$, the $a_i$ component tells us what
to do if the next call event is a handoff of QoS class $j$.  Of course,
if the next call event is a departure, then we again take the obvious
transition.  In \cite{RossTsang89}, it was suggested that this
representation
would lead to computational efficiencies.  However, since we save a
factor of only $3K+1$ in the size of the state space while increasing
the number of possible actions by a factor of $2^K$, and since the
standard algorithm runs in time $|S|^2|A|$, it seems unlikely that this
approach would indeed be more computationally efficient.

\subsection{The Reward Function}
\label{sec-reward_function}
The reward function makes use of the utilities of the QoS classes.
However, these utilities do not completely determine the reward.
For example, if the utility of QoS class $i$ is $u_i$, do we obtain the
utility (i.e., reward) only once (say when the call is connected)?  Do
we obtain it for each unit of time that the call is connected?  The
first possibility corresponds to a charge of a flat rate per call (with
perhaps some penalties for dropping a call or blocking it); the second
corresponds to a charge per call that is a linear function of its
duration.  Clearly other schemes are possible as well.
Our approach can easily accommodate both flat-rate pricing and linear
pricing in a natural way.
We represent the reward $R$ by a
matrix $(R_{ij})$, $i = 1, \ldots, K$, $j=0,1,2$.
Roughly speaking $R_{i0}$ is the reward for accepting a call of
QoS class $i$, if we are thinking of flat-rate pricing, and the reward for
carrying a call of QoS class $i$ for a unit of time, if we are thinking
of linear pricing. $R_{i1}$ is the penalty for blocking
a call of QoS class $i$ and $R_{i2}$ is the  penalty
for dropping a call of QoS class $i$.
With the reward matrix in hand, we can now describe the reward function
$R(s,a)$ in a straightforward way for both flat-rate pricing and linear
pricing.  (We do not consider other reward policies here.)
For flat-rate pricing, we have:
\begin{itemize}
\item $R((\vec{x},\calevts),\accept) = R_{i0}$ if $\calevts = (\arevt,i)$
\item $R((\vec{x},\calevts),\accept) = 0$ if $\calevts$ is $(d,i)$, $(h,i)$, or $n$
\item $R((\vec{x},\calevts),\reject) = R_{i1}$ if $\calevts$ is $(\arevt,i)$
\item $R((\vec{x},\calevts),\reject) = R_{i2}$ if $\calevts$ is  $(h,i)$
\item $R((\vec{x},\calevts),\reject) = 0$ if $\calevts$ is $(d,i)$ or $n$.
\end{itemize}
For linear pricing, we have:
\begin{itemize}
\item $R((\vec{x},\calevts),\accept) = R_{i0} + \sum_{j=1}^K x_j R_{j0}$ if $\calevts$
is $(\arevt,i)$ or $(h,i)$
\item $R((\vec{x},\calevts),\accept) = \sum_{j=1}^K x_j R_{j0}$ if $\calevts$ is
$(d,i)$ or $n$
\item $R((\vec{x},\calevts),\reject) = R_{i1} + \sum_{j=1}^K x_j R_{j0}$ if $\calevts$
is $(\arevt,i)$
\item $R((\vec{x},\calevts),\reject) = R_{i2} + \sum_{j=1}^K x_j R_{j0}$ if $\calevts$
is $(h,i)$
\item $R((\vec{x},\calevts),\reject) = -R_{i0} + \sum_{j=1}^K x_j R_{j0}$ if
$\calevts$ is $(d,i)$
\item $R((\vec{x},\calevts),\reject) = \sum_{j=1}^K x_j R_{j0}$ if
$\calevts$ is $n$.
\end{itemize}

\section{Computing the Optimal Policy}
\label{sec-computing_optimal_policy}
Our goal is to find the policy
(mapping from states to actions) that maximizes
the average sum of rewards. The optimal policy can
be obtained
using dynamic programming, by a modification of
standard techniques like value iteration or policy
iteration (see \cite{Puterman94} for more details).
To explain why we need to modify the standard techniques, we first
briefly review
the value iteration algorithm (although the points we make
apply equally well to policy iteration, the other standard approach, and
all their variants).
The value iteration approach is based
on the following observation.
Let $p_{xy}^a$ be the
probability of making the transition from state $x$ to state $y$ if
action $a$ is taken (this probability is defined by the transition
function).  Suppose $\pi$ is some policy for the MDP.
 Let $V_\pi(x)$ be the value
of state $x \in S$ for policy $\pi$, that is,
the expected reward if we use policy $\pi$ starting in state $x$.
The idea behind value iteration is that we can compute the
optimal policy $\pi^*$ and the optimal value function
$V^*$ by successive approximations.  We start with an arbitrary
value function $V_0$.
Let $\pi_0$ be the optimal choice
of action
with respect to $V_0$; that is
\newcommand{\argmax}{\mbox{argmax}}
\begin{equation}
\pi_{0} (x) = \argmax_{\scriptscriptstyle a \in
A}\left( R(x,a)+\sum_{y \in S} {P_{xy}^{a}}V_0(y)
\right),
\end{equation}
where $P_{xy}^a$ is the probability $T(x,a)(y)$ of making the transition
from $x$ to $y$ if action $a$ is chosen.
Suppose we have defined $\pi_0, \ldots, \pi_n$ and $V_0,
\ldots, V_n$.  We then define $V_{n+1}$ and $\pi_{n+1}$ as follows:
\begin{equation}
V_{n+1}(x) = \max_{\scriptstyle a \in A} \left(
R(x,a)+ \sum_{y \in S} {P_{xy}^{a}} V_{n}(y) \right);
\end{equation}
\begin{equation}
\label{eq:optimal policy}
\pi_{n+1} (x) = \argmax_{\scriptscriptstyle a \in
A}\left( R(x,a)+\sum_{y \in S} {P_{xy}^{a}}V_{n+1}(y)
\right).
\end{equation}
It is a standard result that $\pi_{n+1}$ converges to
an optimal policy $\pi^*$ and $V_n$ converges to the value $V^*$ of
$\pi^*$. 
In practice, we choose some $\epsilon$ and stop the computation when
$|V_n(x) - V_{n+1}(x)| < \epsilon$ for all states $x \in S$; $\pi_n$ is
then an acceptable approximation to $\pi^*$.

We want to apply value iteration to computing optimal admission
decisions.
But, as we hinted above, there may be a problem even defining the MDP.
Admission decisions in cell $c$ must take into account how many
calls will be handed off from neighboring cells. If many more handoff
calls are likely to arrive, the admission decision must make new call
admission decisions more conservatively.
How many handoff calls will arrive
at a cell $c$ depends on the number of calls in $c$'s neighbors and on
the mobility
pattern.  However, if $c$'s state does not include the
number of calls in neighboring cells (as is the case in our model), we
must find some way of estimating the number of calls at $c$'s neighbors
in order to estimate the handoff probability (which in turn affects the
transition probability in the MDP).   There are two (conflicting)
intuitions regarding this estimate.
The first is that the number of calls in the current cell (which is
part of the cell's state) is a good estimate of the number of calls at
its neighbors.  Note that this says that the number of calls in a cell
and its neighbors is strongly correlated.   The second intuition
suggests that, given a policy,
a good estimate of the number of calls
in each QoS class
at a neighboring
cell is just the expected number of calls
in each QoS class
in a state over time as the
policy is run.  
This intuition suggests
that, given a policy,
the number of calls at $c$'s neighbors is uncorrelated with the number
of calls at $c$.

If the first intuition is correct (as we expect it will be in some
cases), then it is relatively straightforward to determine the
probability that a call will be handed off to a cell $c$ in state $s$
(given the mobility model) and thus to define the transition
probabilities for the MDP.  However, if the second intuition is correct
(as experimental evidence shows that it is under the assumptions of our
simulation) then the handoff probability (and hence the transition
probability) depends on the policy.  Thus, it seems we cannot even
define the MDP, let alone use value iteration to compute the optimal
policy!  (We remark that this problem seems not to have been noticed in
other papers that use MDPs in this context \cite{rnt97,TekJab92,YoonUn93}, which simply
assume that the handoff probability is fixed, independent of the policy,
and is given as part of the model.)

Fortunately, even if the second intuition is more appropriate,
it is still often possible to find the optimal policy
by a relatively simple modification of value iteration.
We start by guessing a vector $\vec{c}_0$ that describes the expected
number of calls in each QoS class.
Under the second intuition, provided the guess is correct, it
(along with the known call arrival probability and mobility model)
is enough to determine the handoff probability and thus the transition
probabilities, and hence define an MDP.
We then use standard value iteration to
obtain
the optimal policy $\pi_0^*$
for the resulting MDP.
Under minimal assumptions (namely, that for any two states $s$ and $s'$,
the probability of reaching $s'$ starting in $s$ is positive when using
policy $\pi_0^*$, which is certainly the case for this problem),
the policy $\pi_0^*$ determines a 
{\em stationary probability distribution\/}
$P_0$ over states \cite{Puterman94}.
The probability of a state $s$ according to $P_0$ can
be viewed as the probability of finding a cell in state $s$ if we sample
at random.
Let $\vec{c}_1$ be the expected number of calls of each QoS class
according to $P_0$.  We then use $\vec{c}_1$ to calculate the handoff
probability, and thus determine the transition probabilities for a new
MDP.
We can then calculate the optimal policy $\pi_1^*$
for this MDP.
We then iterate this procedure.
In general, it seems that this approach should
converge under some reasonable assumptions,
but we have not yet proved this analytically.
It does converge for our simulation.
However,
even if it converges, there is no guarantee that it will converge to
an optimal MDP.  All that we can guarantee in general is that it
converges to a local optimum.  
Methods such as simulated annealing \cite{Kirk83},
genetic algorithms \cite{Goldberg89}, or Tabu search \cite{Glover97}
should be useful for finding a global optimum.
\section{Experimental Results}
\label{sec-experiments_with2QoS}
We have compared our MDP approach to one well-known heuristic in the
literature: the NAG policy \cite{ChoiShinMobi98,NagSchw96}.
We have chosen NAG because it is
reported in  \cite{ChoiShinMobi98} to be one of the best
admission-control policies in terms of balancing utilization
efficiency and connection-level QoS.

\subsection{Experimental assumptions}
\label{sec-experimental_assumptions}
In performing our experiments, we made a number of assumptions.
To make the simulations easier to run, we assumed
that there are only two QoS classes, one for data and one for
video.
We used the
 mobility model proposed in \cite{Guerin87},
extended to account for multiple
QoS classes and the finite capacity of the cell BS.
We also made the following traffic and mobility assumptions:
\begin{itemize}
\item The region is covered by $19$ hexagonal cells, each cell with a
radius $R = 1 Km$.
In order to simulate a large area, when a
call is handed off
beyond the region boundary,
it is handed back to another cell
on the boundary (with cells having fewer neighbors having a
proportionately higher chance of being chosen).  This adjusts for the
fact that, otherwise, the number of handoffs received on average at the
boundary cells will be smaller than at the interior cells.
\item Mobiles can travel in any direction with equal probability
and  with a constant speed $SP$.
\item Each mobile goes through changes of direction at ``arbitrary''
places and times.
\item Connection requests are generated according to a Poisson process
with rate $\lambda$ (connections/second/cell) in each cell.
\item Each connection can be a data call with probability $R_{dv}$ or
a video call with probability $1-R_{dv}$.
\item The call-holding time distribution is the same for both video and
data, and is exponentially distributed with mean 120 seconds ( $\frac{1}{\mu} =
120$ seconds).
\item The dwell time is exponentially
distributed with mean $\frac{1}{\rho \mu}$ seconds, where
$\rho=\frac{(3+2\sqrt{3})SP}{9 \mu   R}$.
\item A video call occupies $4$ BU, where a BU is defined to be the
bandwidth to carry out one voice call.
\item Each cell has a fixed capacity of 100 BU.
\end{itemize}

Identical assumptions have been made in many similar studies
\cite{TKChoiDas99,TKNag99,TKDasChoi99,LuBharg96,NagSchw96,SaqYat95,TekJab92,YenRos94,YenRos97,YoonUn93}
so, for the sake of comparison, we make them here as well.
While these assumptions seem quite reasonable in many contexts,
we are well aware that
they may be quite inappropriate in others.
For example, if we are considering traffic along a highway, assuming
that the mobile's movement is random is clearly incorrect, although it
may be more reasonable if we are considering calls in midtown Manhattan.
The assumption that call-holding times are exponentially distributed
becomes less appropriate
as wireless traffic starts to approximate Internet traffic; studies of
the Internet suggest that heavy-tailed distributions are more
appropriate
\cite{PaxFlo95,PaxFlo97,wtsv97}.
Indeed, recent studies of telephone traffic suggest that even for
telephone traffic, heavy-tailed distributions may be appropriate
\cite{dmrw94},
although work by Guerin
\cite{Guerin87}
supports the use of the exponential distribution.
In any case, we stress that we are making these assumptions only for
our case study.
Nothing in our general framework depends on them.

Under the assumptions made here, the modeling approach we used
in Section~\ref{sec-formulating_admission_control} is appropriate.
Our traffic and mobility model do not depend on time. In addition, our
assumptions about Poisson arrival process and exponential call holding
time guarantee the {\em Markov} property, so we do not need to put time
into the state space.  We consider flat rate and linear rate pricing;
therefore, we do not keep track of how long each call has been in the
system. We assume that traffic patterns are homogeneous. The state of the
neighboring cell affects the current cell only through the arrival rate
of handoff calls,
which depends on the number of calls in the cell.
Since experimental evidence shows that, under our assumptions, the
number of calls at a neighbor is the expected number of calls in a state
over time as the policy is run (i.e., the second intuition  holds), we
can compute the handoff probability and the optimal policy by using the
modified value iteration approach discussed in
Section~\ref{sec-computing_optimal_policy}.
In applying this approach, it is necessary to calculate the transition
probabilities for the MDP given an assumption $\vec{c}$ about the
expected number of calls of each class.  As we observed earlier, all
that is needed is to compute the relative probability of the possible types of next call
event; the detailed calculations can be found in
Appendix \ref{appendix-transtionprob}.
We now consider the issue of convergence.
Under the assumptions used in our simulation,
the optimal policy is a threshold \cite{Miller69}.
That is, for
each type of call-arrival event, there is a threshold such that, if
there are fewer calls currently in the system than the threshold, the
call is admitted; otherwise it is rejected.
If there is only one QoS
class, this modification of value iteration converges and computes the
optimal policy.
The reason for this convergence result is as follows:
It is clear that the higher the handoff
probability the lower the acceptance threshold (that is, the lower the
threshold $t$ such that calls are rejected if there are more than $t$
calls already in state $t$), and vice versa.  We can find the optimal
policy by doing a binary search; that is,
start with an assumption $c_0$ about the expected number of calls;
compute the optimal policy
$\pi_0^*$ according to $c_0$, and compute the expected number $c_0'$
of calls corresponding to $\pi_0^*$.
By the argument above, the expected number of calls under the true
optimal policy
is between $c_0$ and $c_0'$.  Let $c_1 =
(c_0 + c_0')/2$.  We compute the optimal policy with respect to
$c_1$, then iterate.  
Although this argument does not work in higher dimensions,
nevertheless, our simulations do always converge.
\subsection{The NAG Policy}
The design goal of NAG is to keep the handoff-dropping probability
$P_{hd}$ below a
certain threshold $\alpha$, thereby maintaining connection-level QoS.
A new call is admitted at cell $c$ if admitting it does not raise the
handoff-dropping probability of existing calls in the system above
$\alpha$ and, if admitted, the new call's handoff-dropping probability
is also below $\alpha$.  The problem is to decide the effect
on $P_{hd}$ of accepting the new call.
NAG does this by estimating
the state of cell $c$ and its neighboring cells $T_{est}$ units of time
after the new-call arrival time,
for some appropriately chosen $T_{est}$. The choice of $T_{est}$ is
based on the current state of each cell and
the aggregate history of handoffs observed in each cell.
\commentout{
{F}rom the assumptions that the behavior of each call is
independent and the probability that a
call is handed off
more than once
during $T_{est}$ is negligible,
it follows from the Central Limit Theorem that
the bandwidth requirements of calls
staying in cell $c$ and calls handing over to $c$ can be approximated
by a Gaussian distribution.
Under our assumptions,
the probability $P_{h}$ that a mobile will
move to
a given neighboring cell
in the next $T_{est}$ time units
and the probability $P_{s}$ that
a mobile will stay in the current cell
during the next $T_{est}$ time units
can be calculated as follows
(see Appendix \ref{appendix-phps}):
\begin{equation}
\label{eq-ph}
P_{h} = (1- e^{-h\mu T_{est}})/6
\end{equation}
\begin{equation}
\label{eq-ps}
P_{s} = e^{-(h+1)\mu T_{est}}
\end{equation}
}%
As observed in \cite{ChoiShinMobi98}, the performance of NAG
depends critically on the choice of the interval $T_{est}$.
It is assumed that $T_{est}$ is
small enough that
the probability that a call experiences a handoff more than once during
$T_{est}$ time units is negligible.
We experimented with different values of $T_{est}$ and chose it to be 5
seconds, since this choice seemed to give the best results for NAG
in our experiments.
\subsection{Numerical Results}
\label{sec-numerical_results}
There are two main issues we want to study in our simulation. First, we
want to understand the behavior of the optimal policy, that is,  what it
depends on and what it does not depend on.  Thus, we study the effect of
user utility, the pricing scheme, and traffic and mobility patterns on the
optimal policy. Second, we want to compare the performance improvements
of the optimal policy over NAG under various conditions. The comparison
will yield insight into the behavior of NAG and provide guidelines for
the design of more efficient heuristic admission-control policies.

For these studies, we set the reward matrix $R_{ij}$
(see Section \ref{sec-formulating_admission_control})
as follows.
We took $R_{i0}$ to be proportional to the
bandwidth
requirement of each QoS class; we set $R_{i1}$ to be negative and its
absolute value to be $10$\% of $R_{i0}$. We varied the ratio
$r_{{db}_{i}}$ 
($R_{i2}$ divided by $R_{i1}$)
to study its effects on the optimal policy.
We considered two values for $R_{dv}$:  $1$ (which means that all calls
are data calls) and $0.5$ (which means that half the calls are data and
half are video).
We also considered two settings for the average speed of the mobile:
100 Km/hr and 50Km/hr.
We refer to the
former setting as the high user mobility case and the latter as the lower
user mobility case.
\subsubsection{The
Characteristics of the Optimal Admission Policy
}
\label{sec-r_db}

\input{epsf}
\begin{figure*}[ht]
\setlength\tabcolsep{0.1pt}
\begin{center}
\begin{tabular}{cccc}
\epsfysize=4.2cm \epsffile{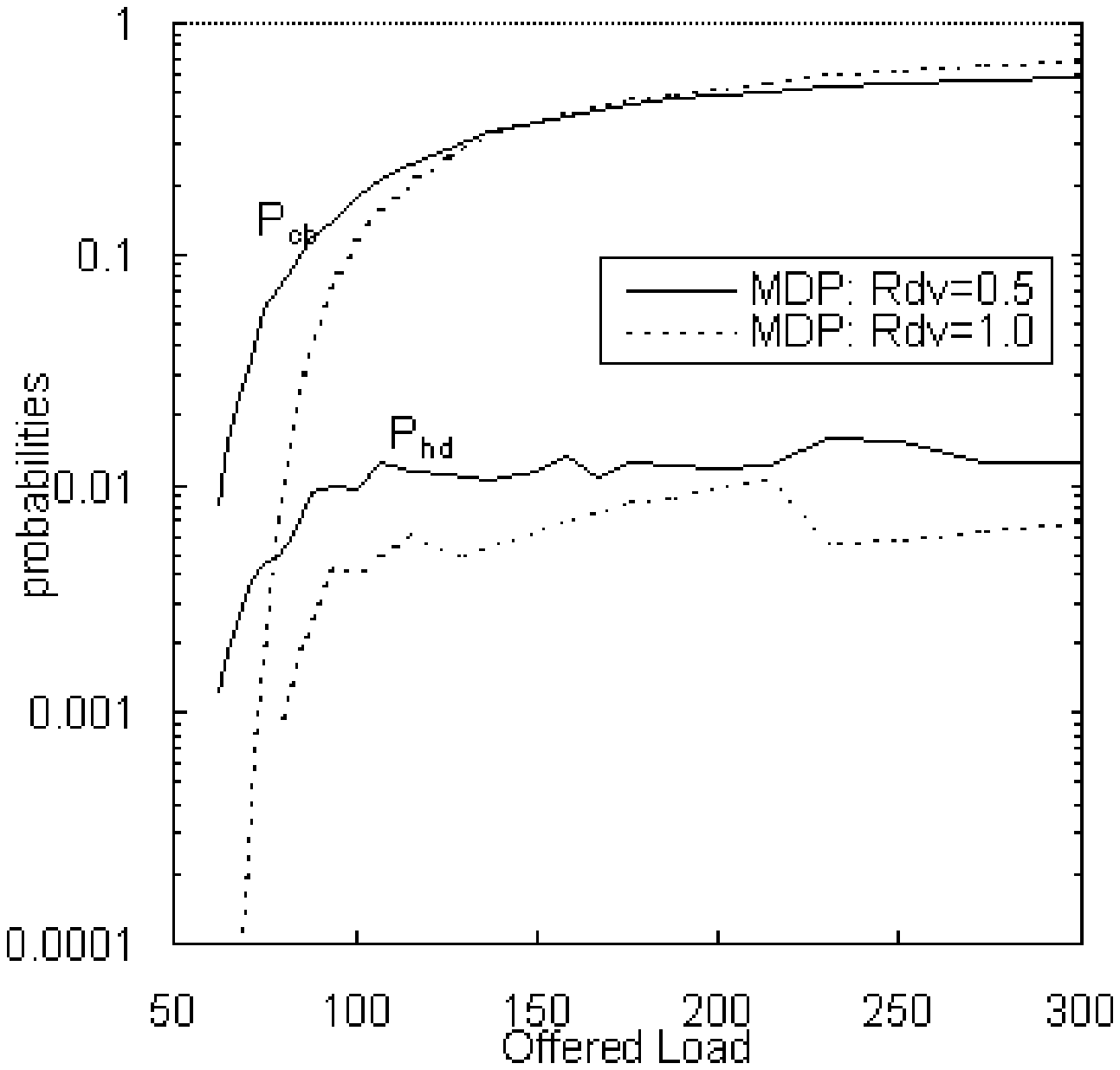} &
\epsfysize=4.2cm \epsffile{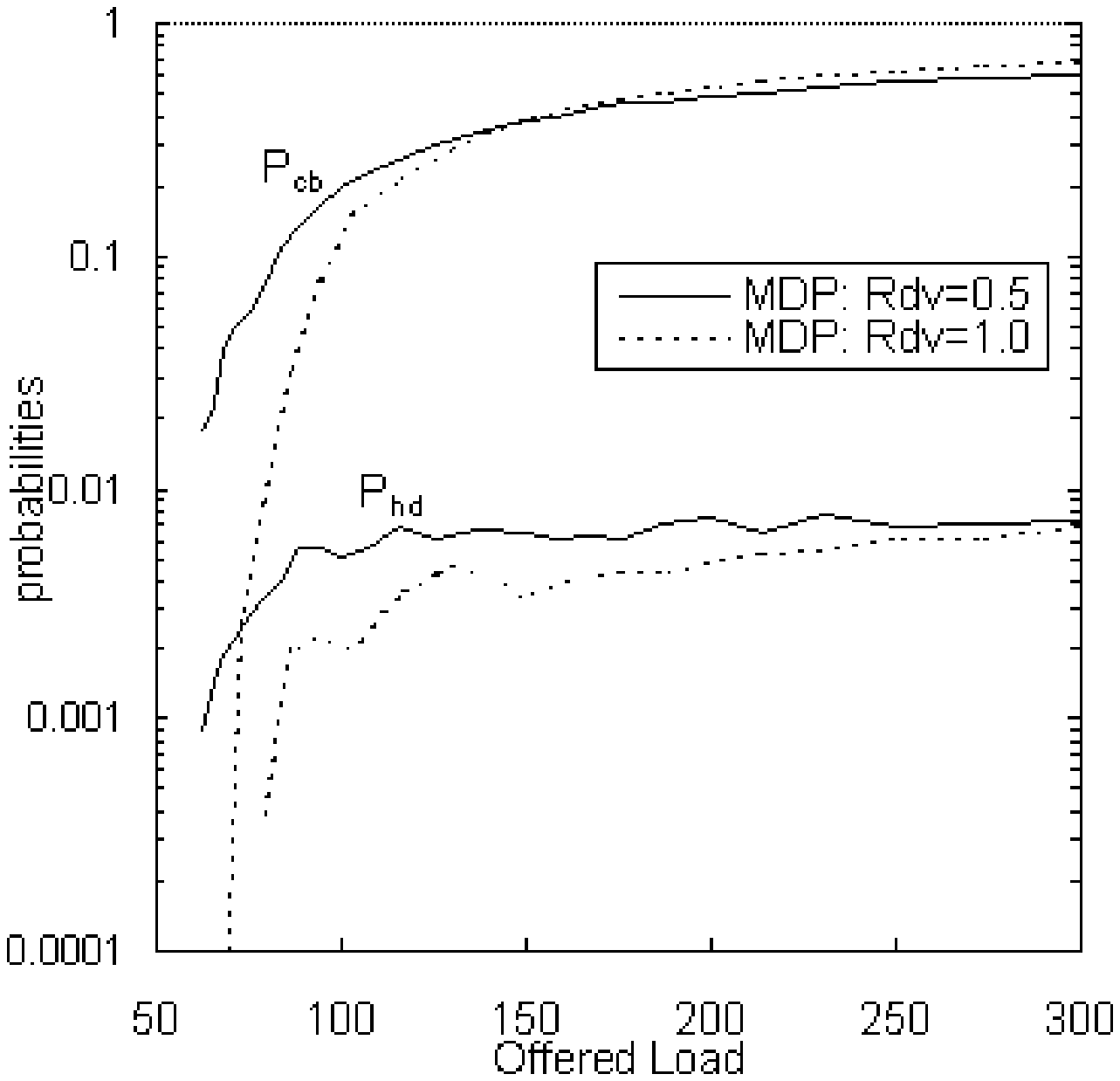} &
\epsfysize=4.2cm \epsffile{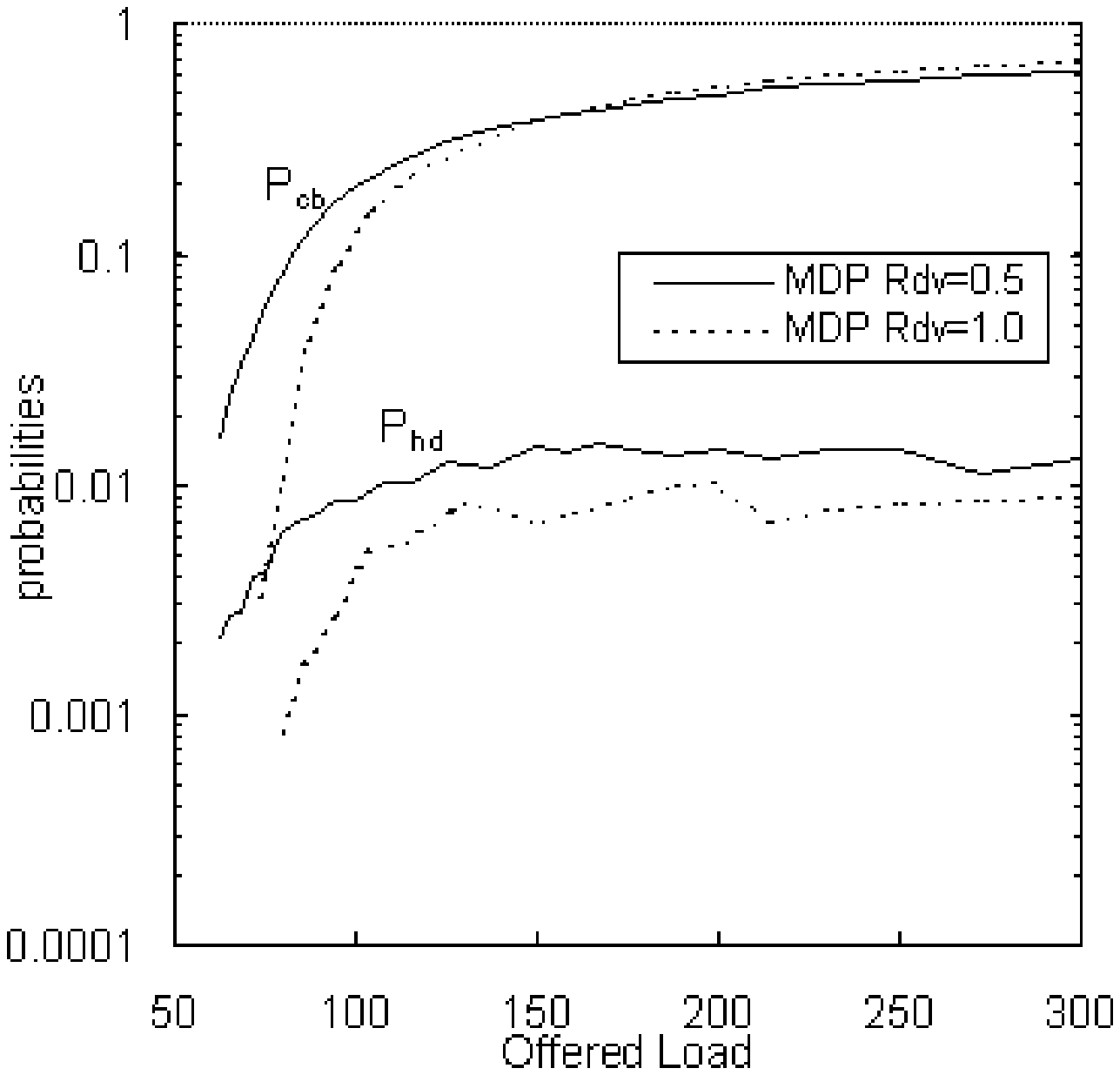} &
\epsfysize=4.2cm \epsffile{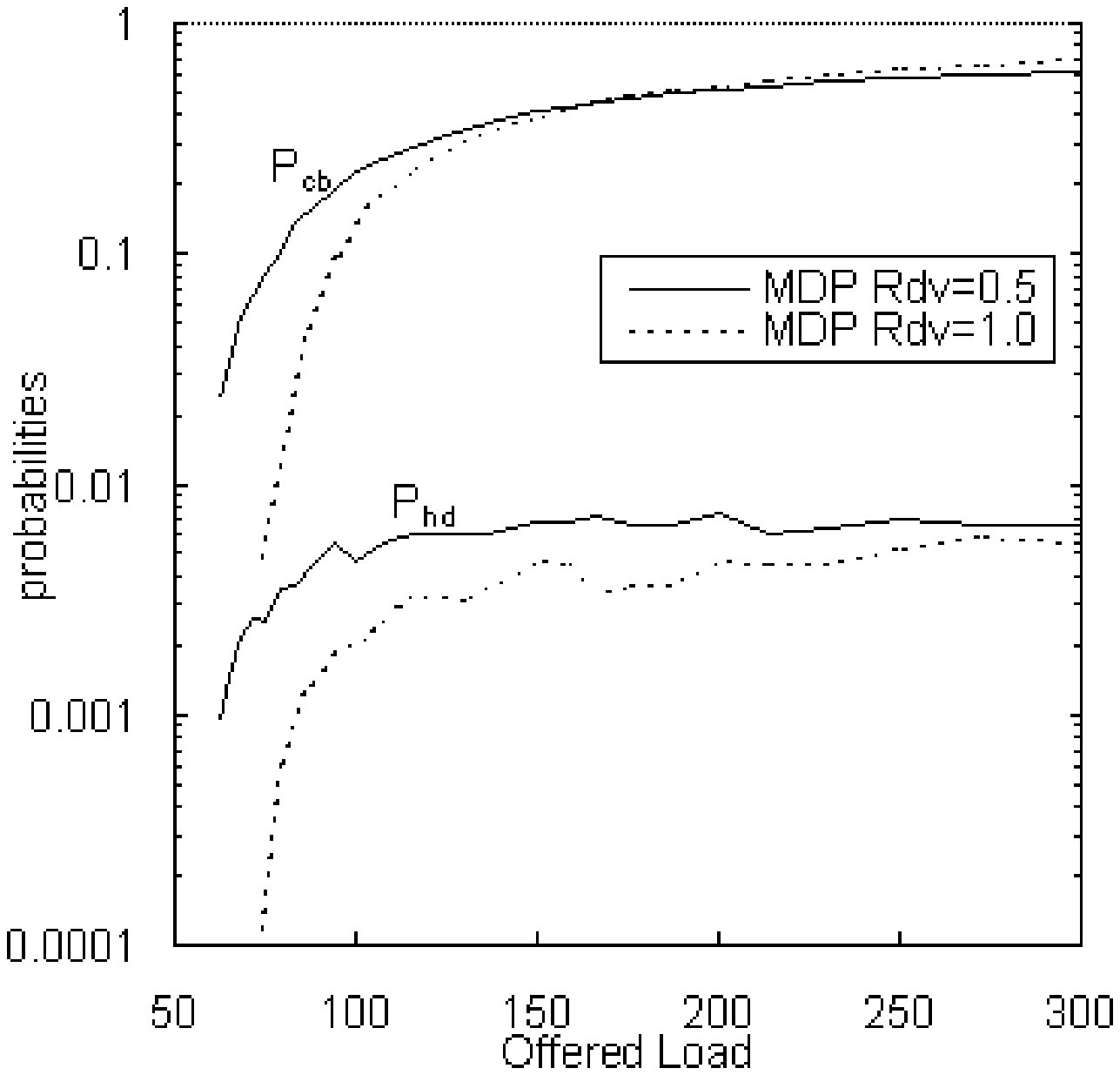} \\
{\footnotesize (a)SP=50: $r_{{db}_1}=r_{{db}_2}=40$}  &
{\footnotesize (b)SP=50: $r_{{db}_1}=r_{{db}_2}=80$}  &
{\footnotesize (c)SP=100: $r_{{db}_1}=r_{{db}_2}=40$}  &
{\footnotesize (d)SP=100: $r_{{db}_1}=r_{{db}_2}=80$}  \\
\end{tabular}
\end{center}
\caption{
$P_{cb}$ and $P_{hd}$ vs.~offered load: optimal admission-control
policy (flat pricing scheme).
\label{MDP_r40_and_80}
}
\end{figure*}
\input{epsf}
\begin{figure*}[ht]
\begin{center}
\epsfysize=4.0cm \epsffile{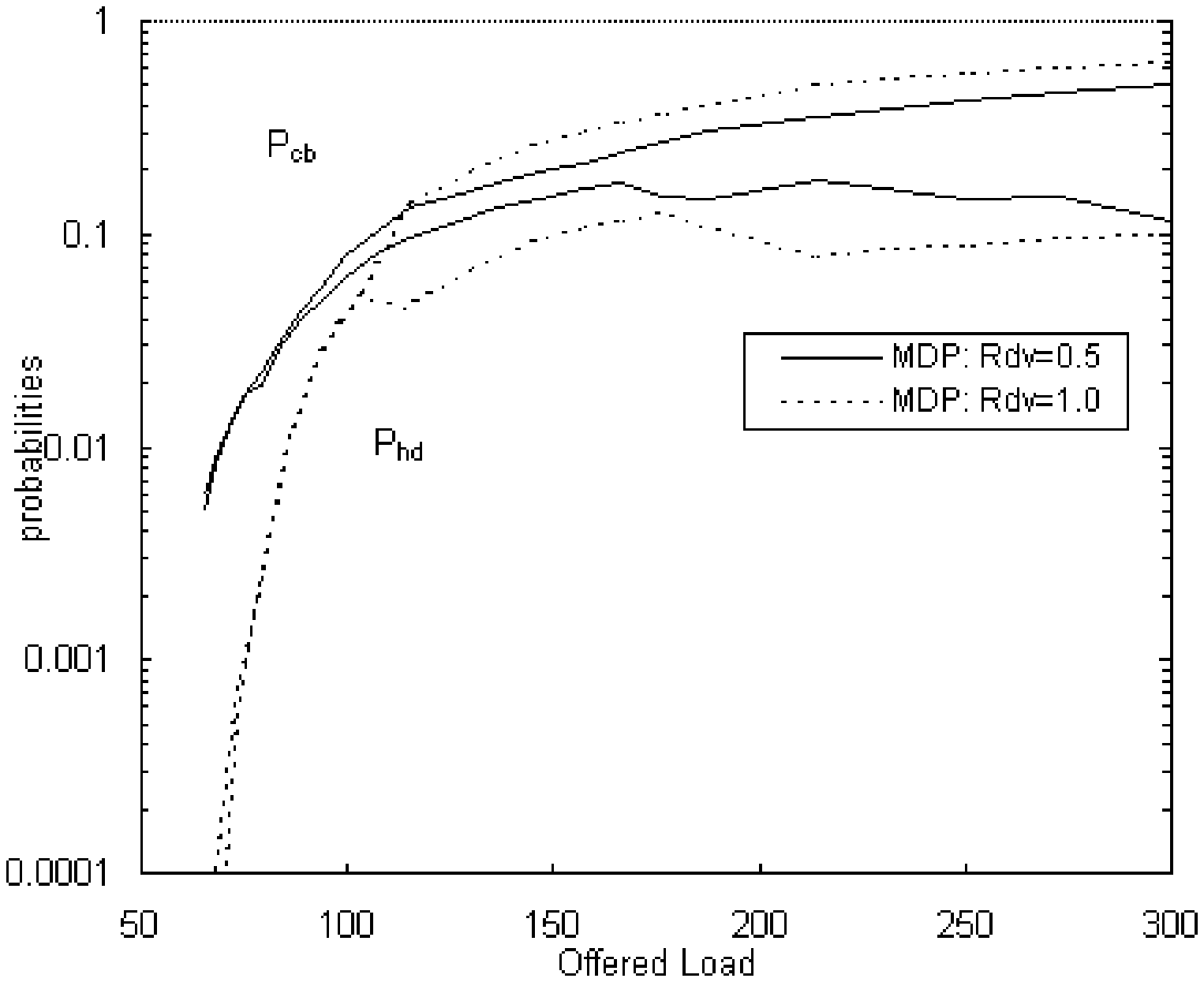}\\
$r_{{db}_1}=r_{{db}_2}=80$
\end{center}
\caption{
$P_{cb}$ and $P_{hd}$ vs.~offered load: optimal admission-control
policy (linear pricing scheme).
\label{MDP_r80_lp}
}
\end{figure*}

First, we want to study the effect of user utility on the optimal
admission policy. To do so,
we vary  $r_{{db}_i}$ (the ratio between call dropping and call blocking
penalty) for the two QoS classes.
In Figure~\ref{MDP_r40_and_80}, we see that for the same offered load,
the call-dropping probability in the case of $r_{{db}_1}=r_{{db}_2}=80$ is smaller than
the call-dropping probability in the case of $r_{{db}_1}=r_{{db}_2}=40$ case.
This is what we would expect, since the
optimal admission-control policy takes the application utilities into
account.
The greater the relative penalty for dropping a call, the lower the
probability that it will be dropped.
Figure \ref{MDP_r40_and_80} also shows that the
optimal admission policy does not depend strongly on the traffic pattern
parameter $R_{dv}$
or on the 
mobility parameter $SP$.

Our simulations do show that the pricing scheme 
can have a significant impact on the optimal control
policy. If $r_{{db}_1}=r_{{db}_2}=80$, the call
dropping probability for the flat pricing scheme is around $1$\%
under  high load
(see Figure \ref{MDP_r40_and_80}(b)),
but for the linear pricing scheme, the call dropping
probability is around $10$\% under the same high load
(see Figure \ref{MDP_r80_lp})
. This result is
also very intuitive. Since the reward ($R_{i0} \times 1/ \mu $) for
accepting a call in the linear pricing scheme is much higher than the
reward ($R_{i0}$) for accepting a call in the flat pricing scheme, the
linear pricing scheme tends to accept more calls than the flat pricing
scheme. 
\subsubsection{The Expected Utility of Different Admission Control
Policies}
\label{sec-utility_gain_experiment}

\begin{figure*}[ht]
\setlength\tabcolsep{0.2pt}
\begin{center}
\begin{tabular}{cccc}
\epsfysize=4.2cm \epsffile{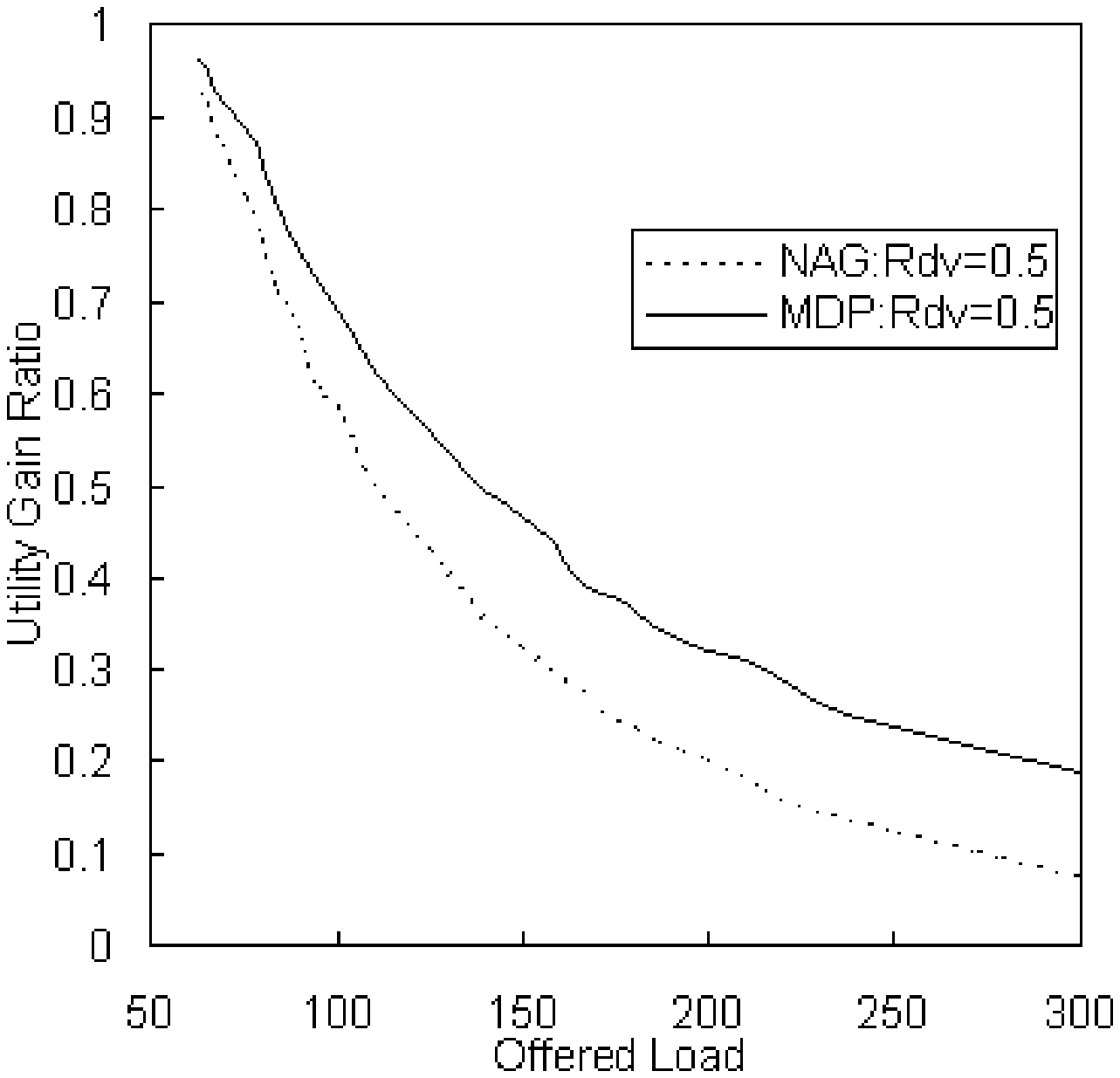} &
\epsfysize=4.2cm \epsffile{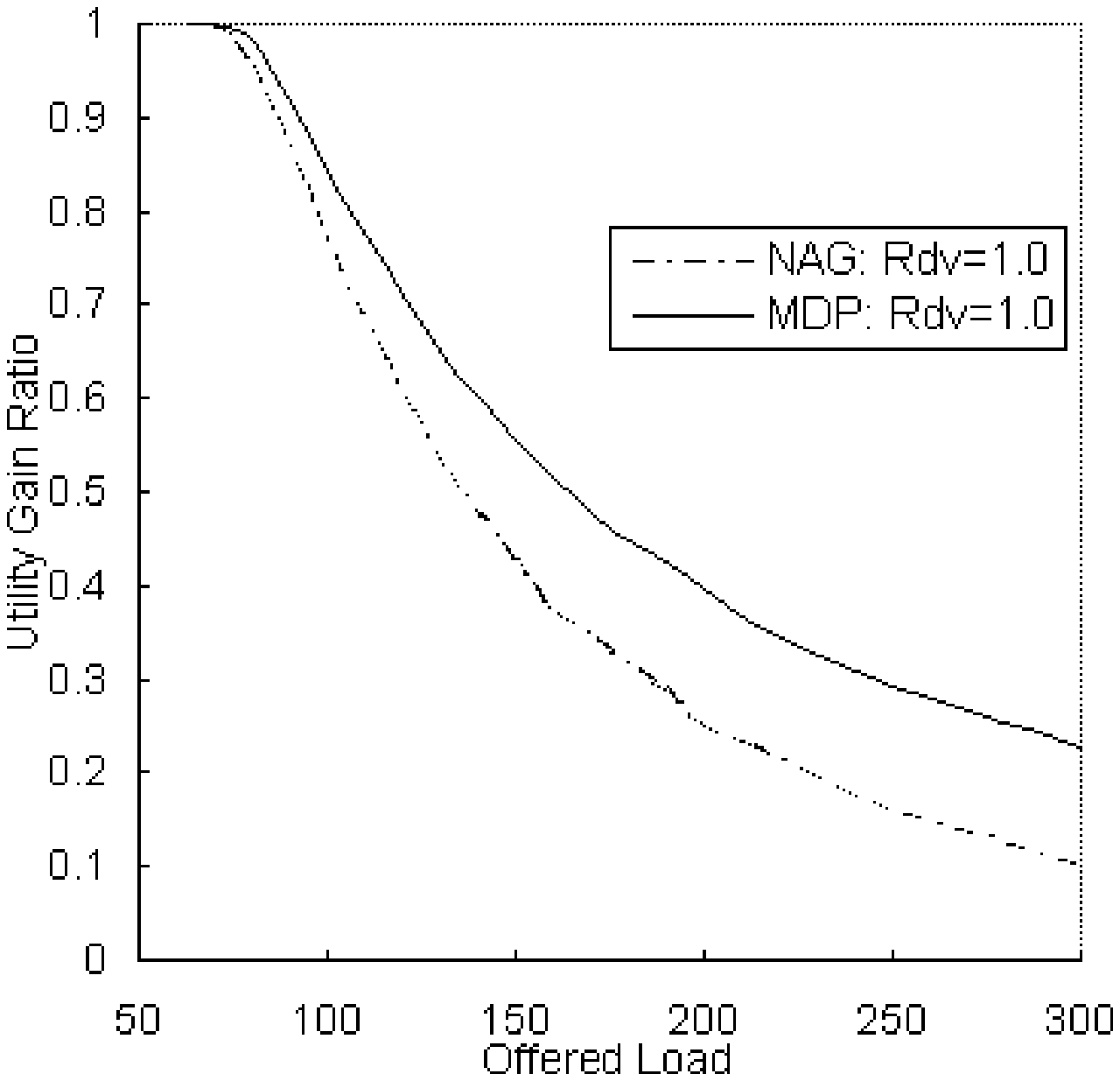} &
\epsfysize=4.2cm \epsffile{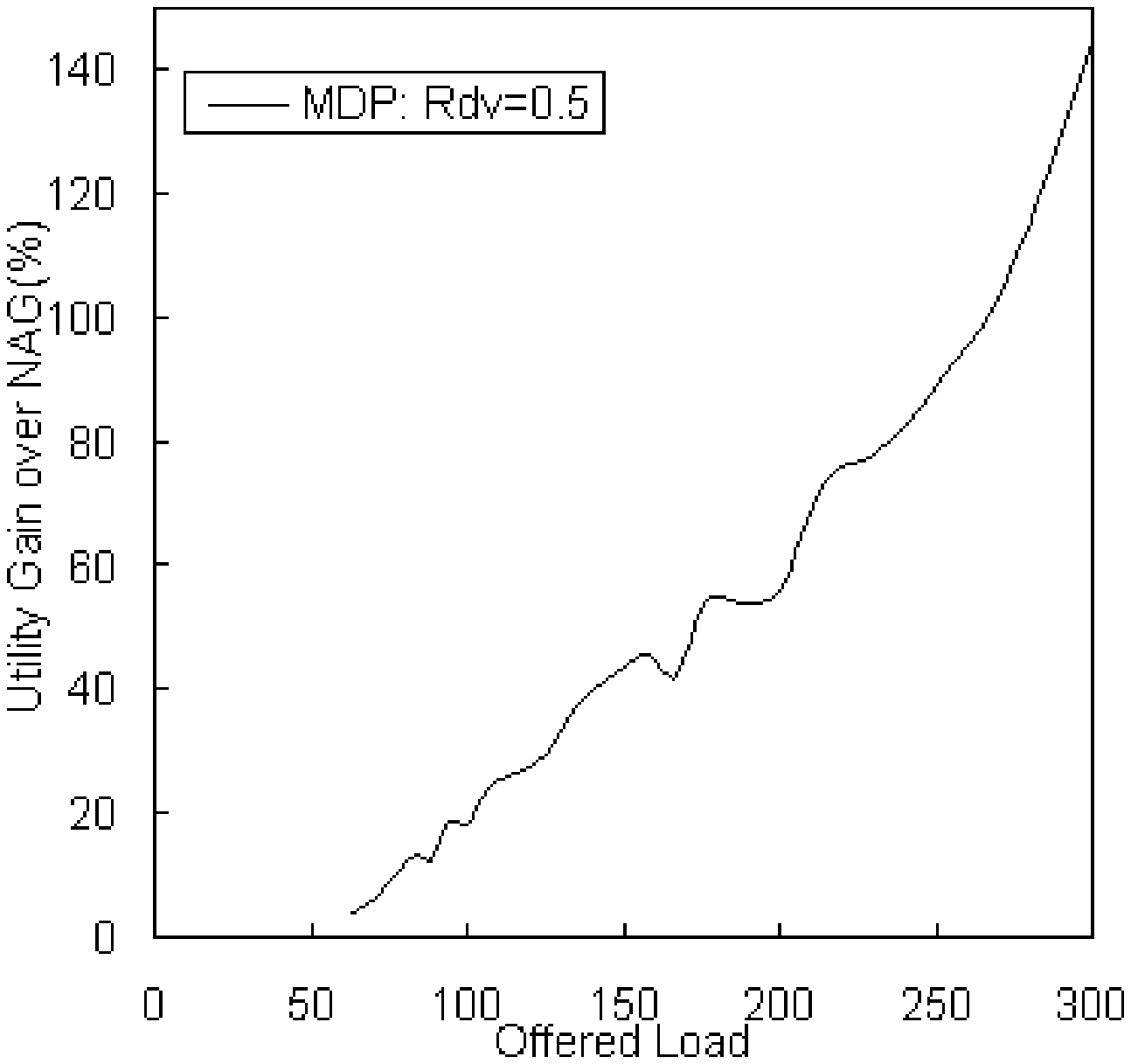}  &
\epsfysize=4.2cm \epsffile{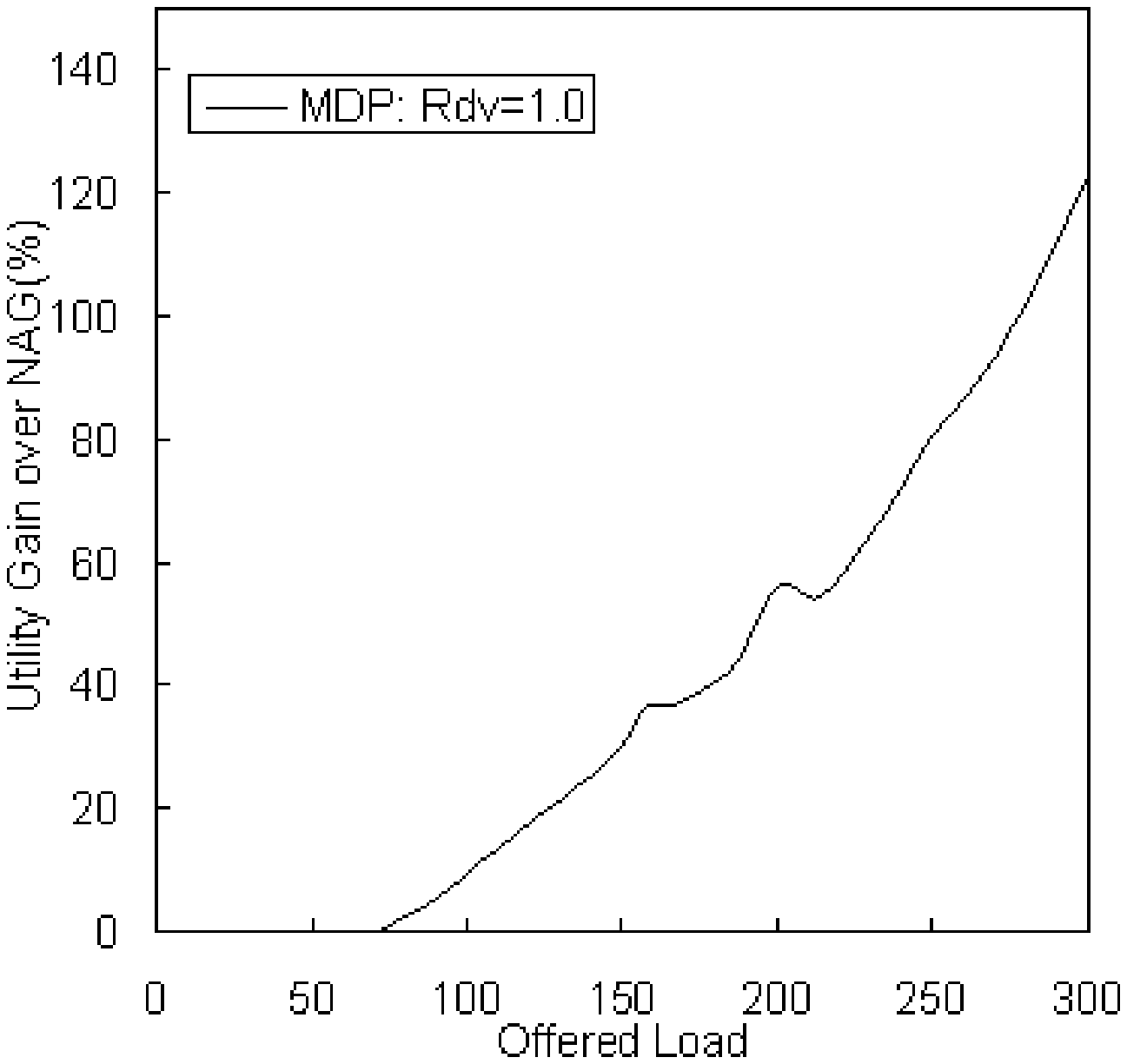}  \\
{\footnotesize (a)  $R_{dv}=0.5$} &
{\footnotesize (b)  $R_{dv}=1.0$} &
{\footnotesize (c)  $R_{dv}=0.5$} &
{\footnotesize (d)  $R_{dv}=1.0$}   \\
\end{tabular}
\end{center}
\caption{
$r_{{db}_1}=r_{{db}_2}=80$, NAG's design goal is $P_{hd}=4\%$. (a),(b):
Expected utility ratio of MDP and NAG; (c),(d): MDP's utility gain
over NAG. 
\label{UGR_r80_phd4}
}
\end{figure*}

\begin{figure*}[ht]
\setlength\tabcolsep{0.2pt}
\begin{center}           
\begin{tabular}{cccc}
\epsfysize=4.2cm \epsffile{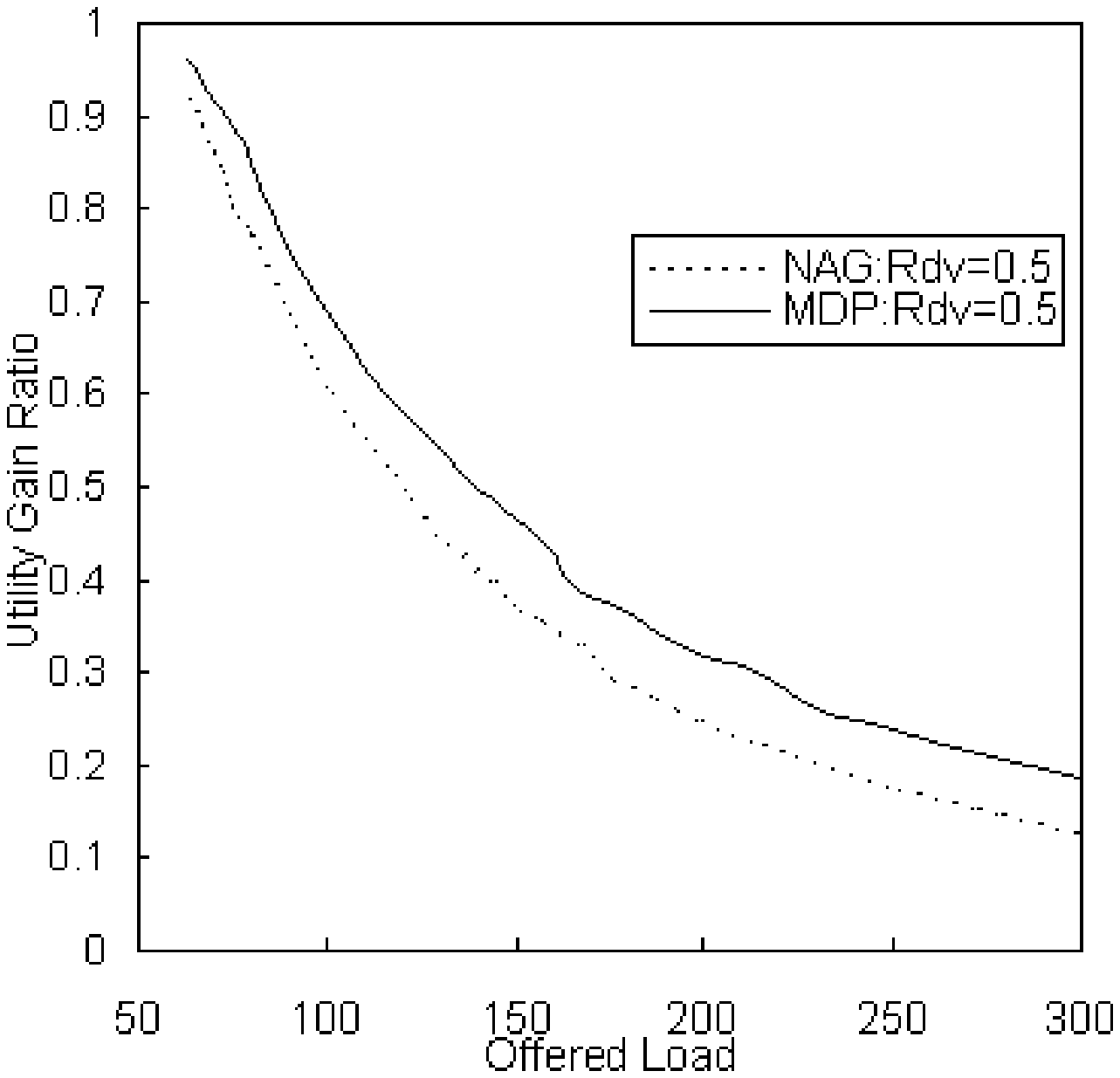} &
\epsfysize=4.2cm \epsffile{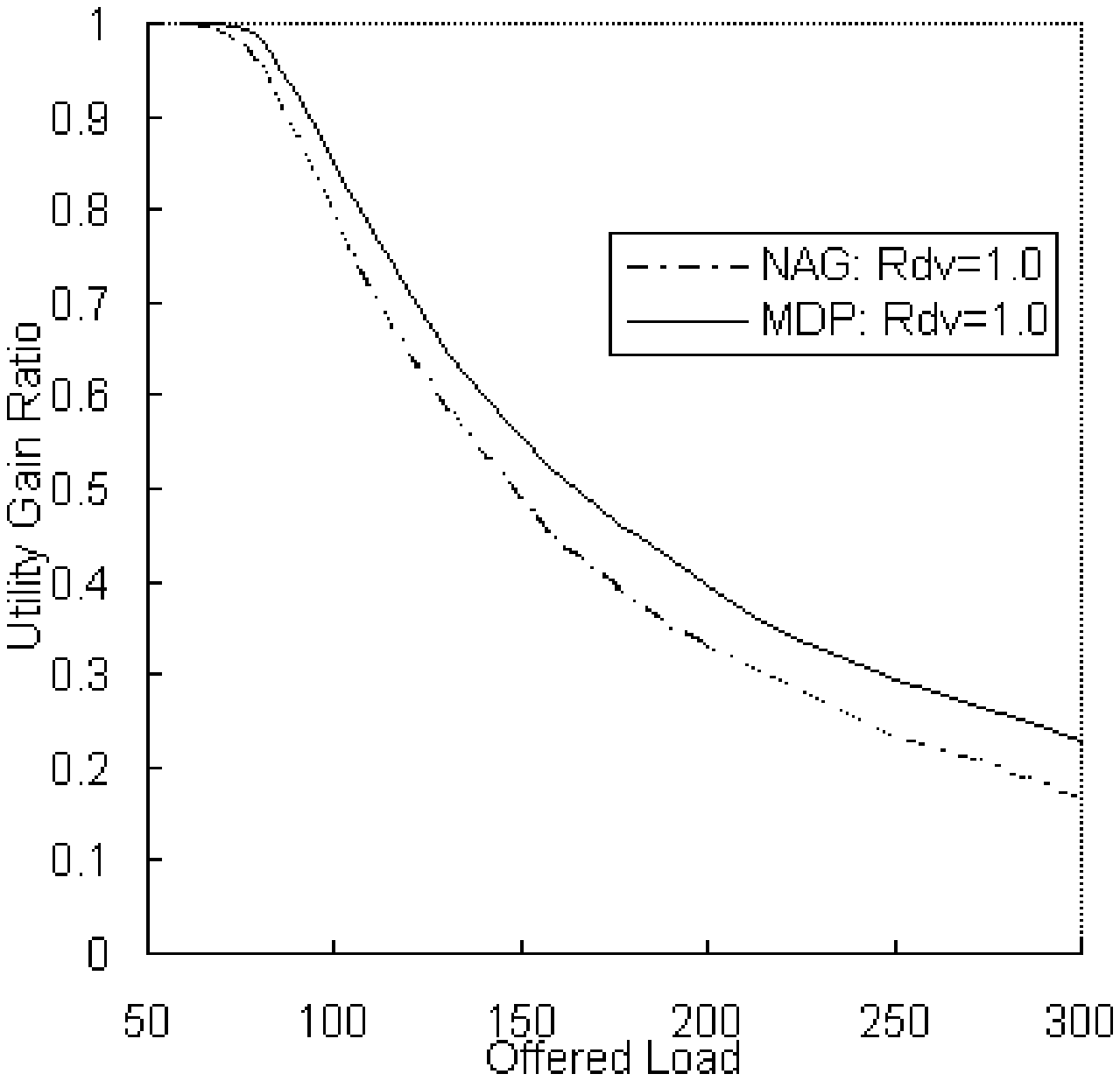} &
\epsfysize=4.2cm \epsffile{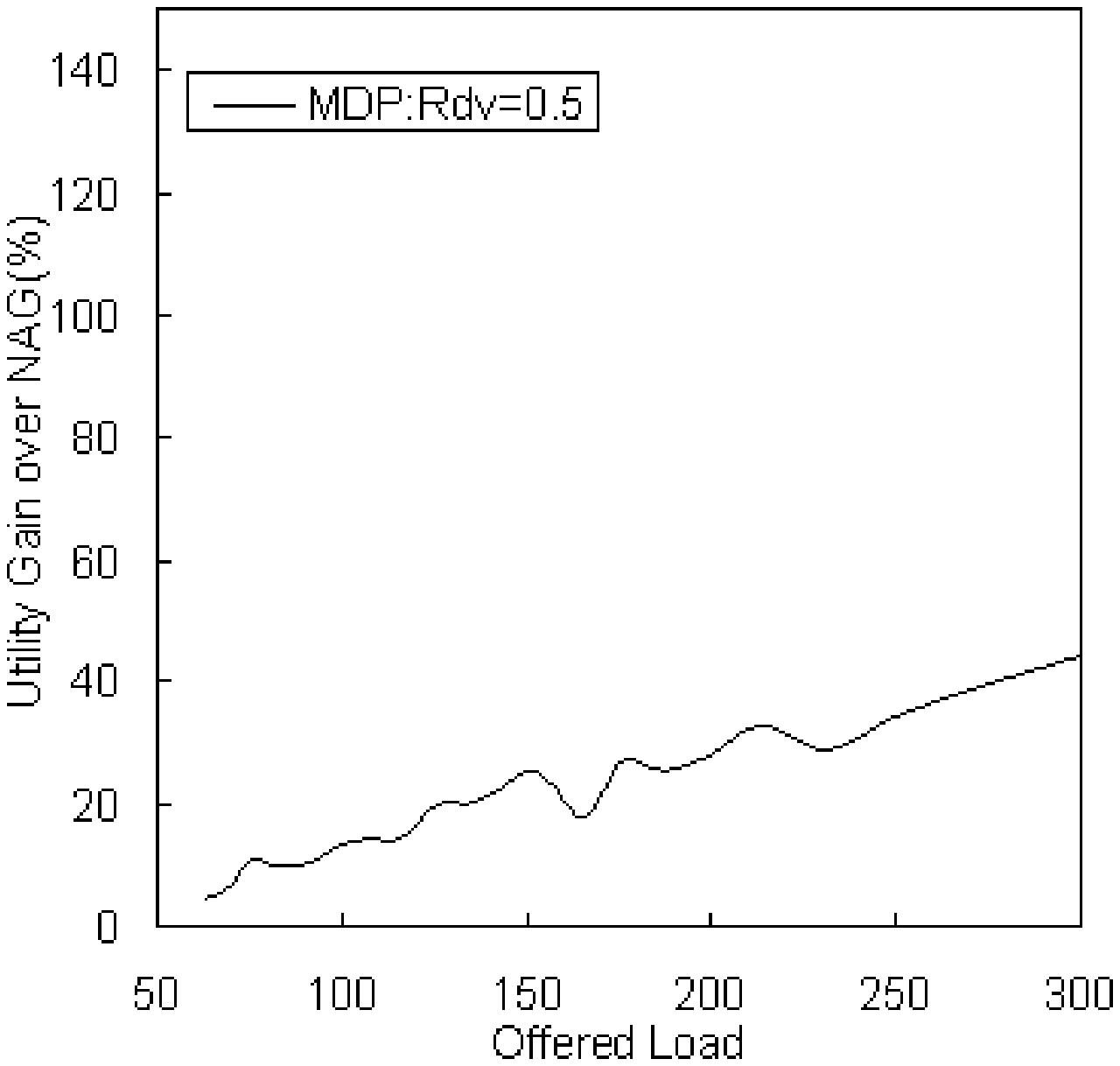} &
\epsfysize=4.2cm \epsffile{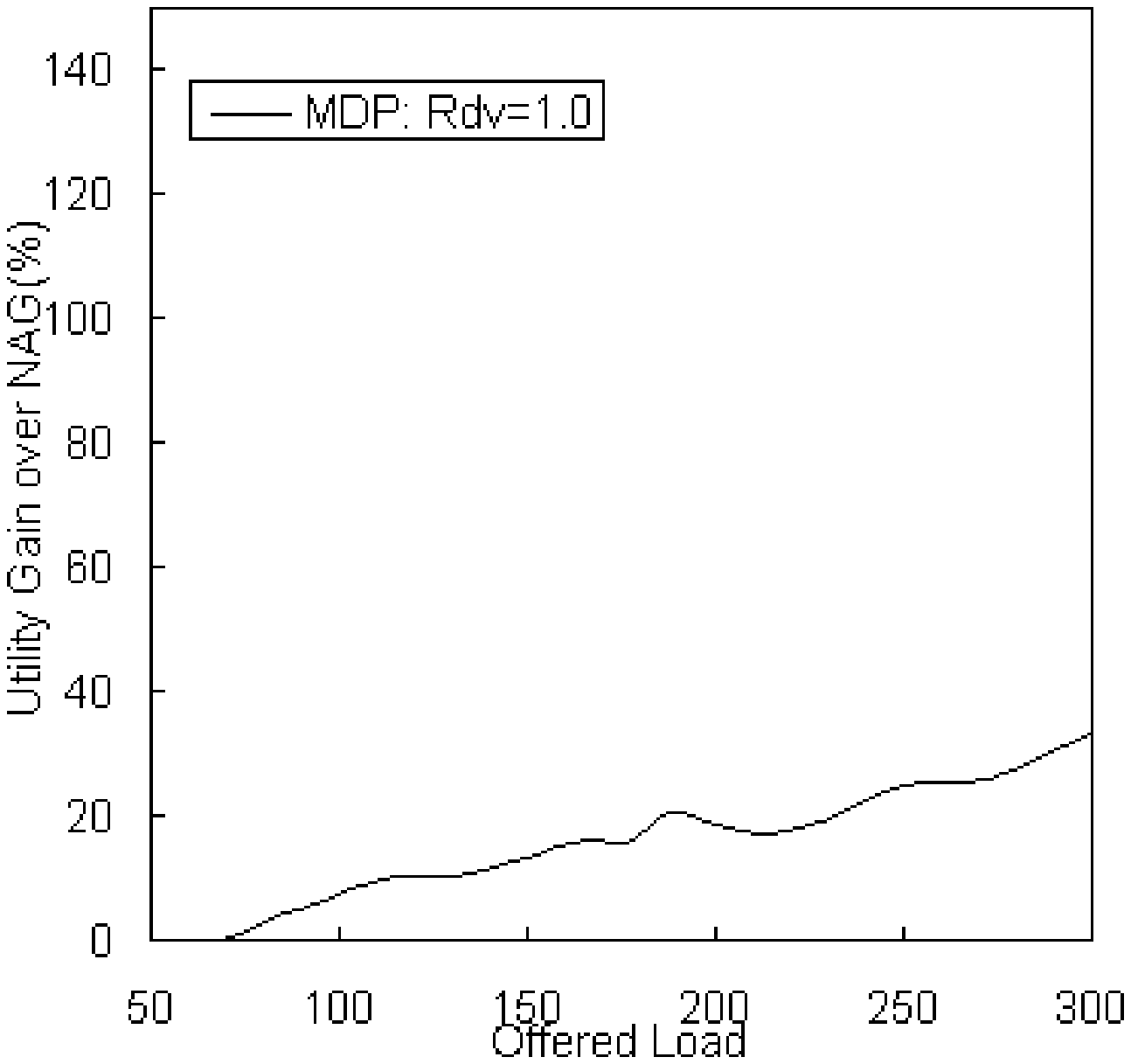}  \\
{\footnotesize (a)  $R_{dv}=0.5$} &
{\footnotesize (b)  $R_{dv}=1.0$} &
{\footnotesize (c)  $R_{dv}=0.5$} &
{\footnotesize (d)  $R_{dv}=1.0$}    \\
\end{tabular}
\end{center}
\caption{
$r_{{db}_1}=r_{{db}_2}=80$, NAG's design goal is $P_{hd}=1\%$. (a),(b):
Expected utility ratio of MDP and NAG; (c),(d): MDP's utility gain
over NAG.
\label{UGR_r80_phd1}
}
\end{figure*}

\begin{figure*}[ht]
\begin{center}
\begin{tabular}{cc}
\epsfysize=4.2cm \epsffile{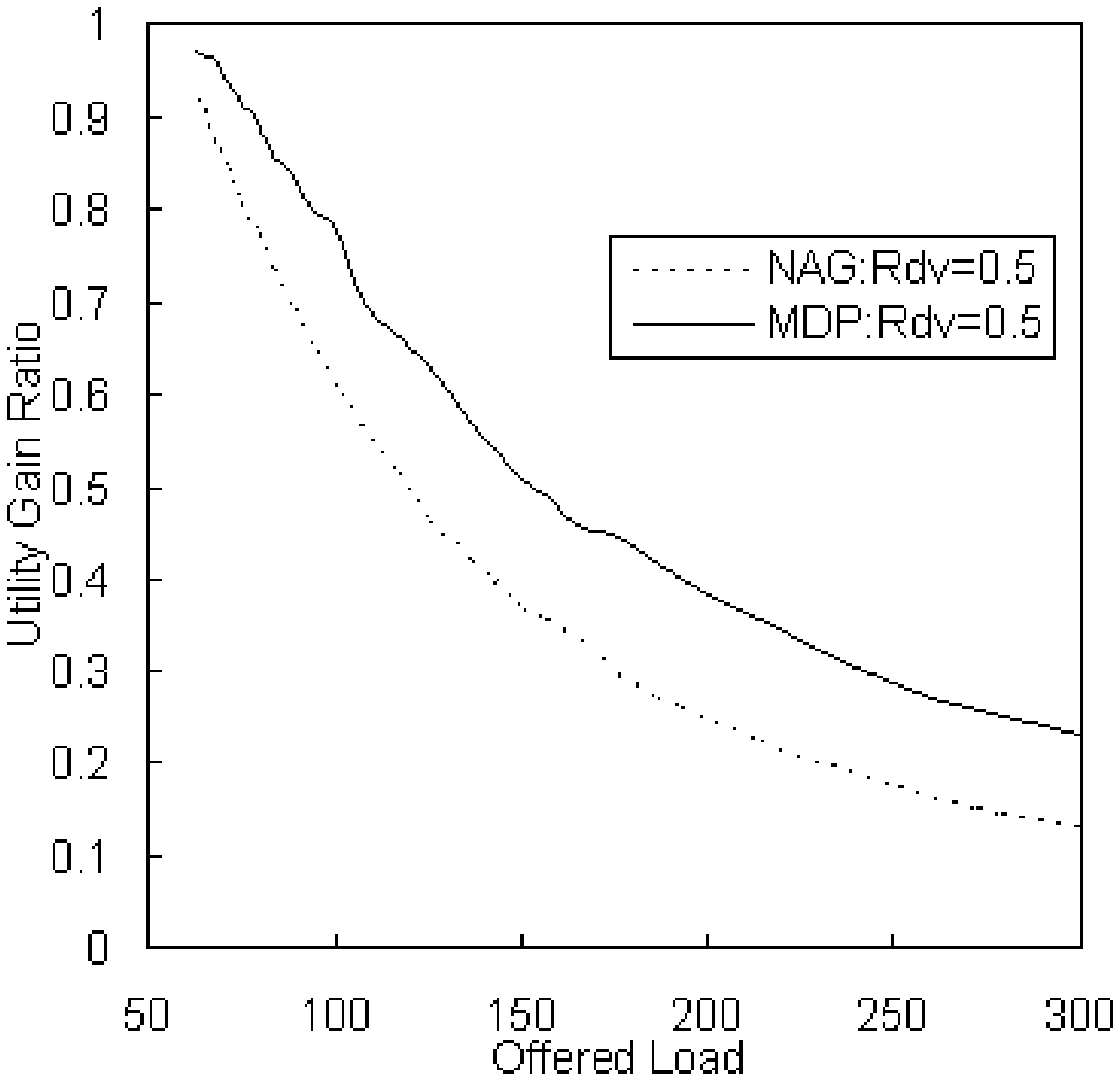} &
\epsfysize=4.2cm \epsffile{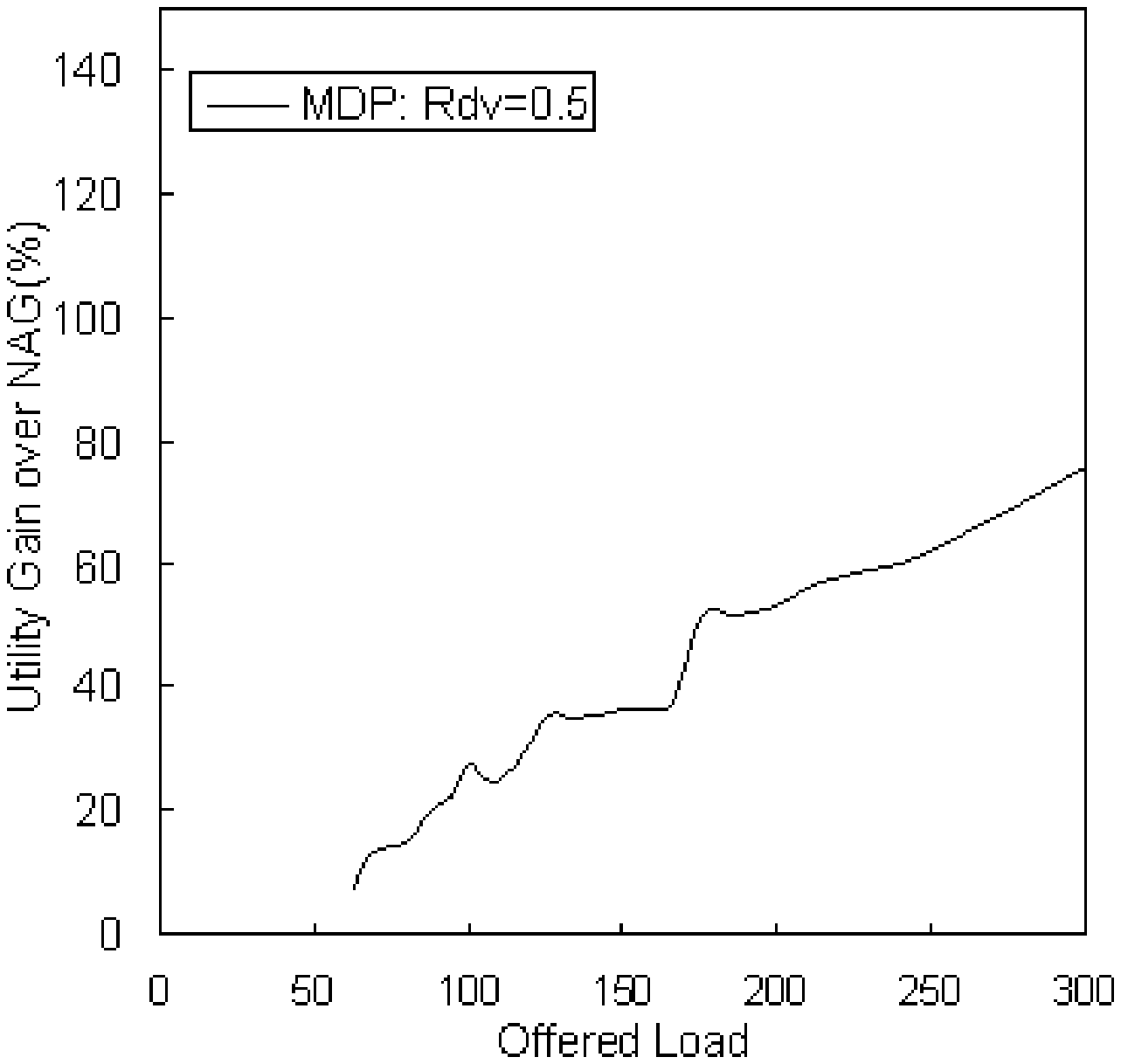} \\
{\footnotesize (a)  $R_{dv}=0.5$} &
{\footnotesize (b)  $R_{dv}=0.5$}    \\
\end{tabular}
\end{center}
\caption{
$r_{{db}_1}=2$, $r_{{db}_2}=80$, NAG's design goal is $P_{hd}=1\%$.
(a): Expected utility ratio of MDP and NAG; (b): MDP's utility gain
over NAG.
\label{UGR_r80_phd1_rdb1eq2}
}
\end{figure*}

The goal of the network service provider is to maximize its revenue
while providing satisfactory service to the users over time. So the
network performance should be measured in terms of the average system
revenue over time.
Since the system's utility reflects the utility of users (to some
extent),
the average system revenue over time also reflects how good is the service
that the system provides to the users over time. In our simulation,
we compare the expected utilities of NAG and the optimal control policy
under different offered loads. To capture the performance degradation of
the network as the load increases, the expected utility is normalized with
respect to a network with infinite capacity. In this section, we
present the results for the flat pricing scheme and lower mobility case.
Similar results hold for the linear pricing scheme and for high
mobility;  we omit details here.

Note that
our optimal admission-control policy takes the user's utility into
account automatically.
However, NAG
does not.  The call-dropping probability
is selected
by NAG in an {\em ad hoc\/} way.
With the right choice of parameters,
the performances gap between our optimal admission-control policy and NAG
decreases.
On the other hand,
with the wrong choice, the opposite is true.
As our experimental results show, there is no one setting of the
call-dropping probability that gives good results over the whole range
of interest.
For example,
Figure~\ref{UGR_r80_phd4}
describes the results
when NAG
picks $P_{hd}=4\%$ as the design goal.
The optimal admission-control policy achieves a performance improvement
of approximately
$18$\%, $55$\% and $144$\% over NAG at offered loads of  $100$, $200$,
and $300$,
respectively, in the $R_{dv}=0.5$ case 
(see Figure \ref{UGR_r80_phd4}(c)). 
(At load $300$, the
improvement of
$144$\% corresponds to
a factor of more than 3 in the call-dropping
probability while
maintaining almost the same call blocking probability.)
With $P_{hd}=1\%$ as NAG's design goal,
as shown in Figure \ref{UGR_r80_phd1},
the optimal admission-control policy achieves a performance improvement 
of approximately
$14$\%, $21$\%, and $40$\%
over NAG at offered loads of $100$, $200$,
and $300$,
respectively, in the
$R_{dv}=0.5$ case.
The improvement in the normalized expected utility
depends
in part on the type of calls that must be dealt with.
For example, the improvements in terms of
normalized expected utility over NAG at offered
load $100$, $200$, and $300$ is $11\%$, $44\%$, and $122\%$, respectively,
in the case of
$R_{dv}=1.0$ when $P_{hd}=4\%$ is the design goal 
(see Figure \ref{UGR_r80_phd4}(d)).
This
is because the optimal policy is biased towards data calls, since data
calls can still be accepted in states that video call cannot be
accepted due to
insufficient bandwidth.
The differences between NAG and our optimal MDP policy become even more
dramatic if we change $r_{{db}_1}$ from 80 to 2 (making the penalty for
dropping a data call much closer to that of blocking a call; as we
observed in the introduction, this might make sense for FTP transfers).
In this case, the improvement of the MDP approach over the NAG approach
increases from 40\% to 70\%
at offered load $300$, with $P_{hd}=1$\% as NAG's design goal
(see Figure \ref{UGR_r80_phd1_rdb1eq2}).

Of course, once we add utilities into the picture,
it would not be difficult to modify NAG so that it attempted
at all times to adjust $P_{hd}$ to obtain optimal performance.
But by doing this, NAG would essentially be approximating our MDP
approach.  In fact, as our state space grows, we will need to find
useful approximation to the MDP approach, since it becomes
computationally too expensive.  NAG may be one such approach, but there
may be better ones as well.  The MDP approach at least gives us a framework by
which to judge this issue.

\section{Discussion and Related Work}
\label{sec-related_work}
There has been a great deal of related work on admission control, both
in the context of wireless and wired networks.
Optimal call-admission policies in cellular networks
which carry only voice calls were studied in~\cite{rnt97,SaqYat95}.
Near-optimal admission-control policies 
\cite{YenRos94,YenRos97} and heuristic approaches 
\cite{ChoiShinSIG98,NagSchw96} have been studied.
Bandwidth adaptation using layered encoding has been studied recently in
\cite{TKNag99,TKChoiDas99,TKDasChoi99}.
None of these papers tried to provide a general decision-theoretic
framework for the call-admission problem.  Rather, they typically
focused on the standard scenario (Poisson arrival times; exponentially
distributed call holding and dwell times) and tried to solve the problem
in that setting.
While we believe that our main contribution is the general
framework, not its application to the standard scenario,
as we saw, our approach does dominate 
a well-known heuristic approach in this setting.

Previous work
has either had the goal of keeping the call-dropping probability below a
given threshold
\cite{ChoiShinSIG98,NagSchw96} or
used a
weighted factor between dropping and blocking as the design goal
\cite{rnt97,SaqYat95,YenRos94,YenRos97}.
The problems with choosing an appropriate handoff-dropping probability
threshold have already been discussed in detail.  The second approach
can be viewed as a special case of ours.
In \cite{rnt97,SaqYat95,YenRos94,YenRos97}
the weighting factor is determined by the relative
utilities of call blocking and call dropping.  Thus, 
these approaches are essentially
a special case of ours, using flat-rate pricing, with only
one QoS class.  They assume that the call handoff rates are fixed, and
do not deal with the fact that they are influenced by the policy.
In addition, these papers do not address subtleties
such as the effect of the policy chosen on the handoff rates.

There is another large body of literature on admission control in
wired networks.   The measurement-based
admission-control (MBAC) approaches
\cite{GrosTse97,GrosTse99,jdsz97,gkt95,TseGros97}.  Of these,
\cite{gkt95} is perhaps most relevant to our work.
For example, in \cite{gkt95},
Renegotiated Constant Bit Rate (RCBR) Service is proposed
for variable bit rate video. The main idea to increase the
throughput through statistical multiplexing
while providing delay and loss guarantees
through renegotiation.  While the basic idea of renegotiation may prove
applicable in the wireless setting, we cannot immediately use this
approach due to significant differences between
the wired and wireless settings.  In particular,
since a wireless channel can be used exclusively by only one flow
at a time,
we cannot multiplex many flows on a wireless channel as easily as 
we can multiplex flows on bandwidth in the wired case.
(For example, if both
flows transmit at the same time, the data will be garbled in the
wireless case.)   
Moreover, in MBAC, even if renegotiation fails, the flow keeps its
original bandwidth. But for the wireless counterpart, if a handoff
fails, the flow loses its original bandwidth.

Our general approach leads to some interesting research issues.
All our algorithms are polynomial in the size of
the state space.
In our model,
we used a number of techniques to limit the size of the state space.
However, in realistic cases, we can expect to have much larger state
spaces.
Although the computation of the optimal solution is done off-line, and the
admission decisions are made simply through table lookup,
there is a limit to the number of states we can feasibly handle.
We are currently exploring techniques to provide good approximate
solutions in the presence of large state spaces. 
Motivated by the result in \cite{KearnManNg99} that the running time
in computing the near-optimal policies in arbitrarily large,
unstructured MDPs does not depend on the number of states, we believe
that MDP can be quite applicable in practice. 
We expect that, if the state space is large (due to many QoS
classes) the behavior of the system will be qualitatively similar to the
case when it is small.

{\small
\bibliographystyle{plain}
\bibliography{lilisbib}
}

\newcommand{\ho}{\overline{\mbox{\it ho}}}

\pagebreak
\appendix

\commentout{
\section{Deriving the Arrival Rate of Handoff Calls}
\label{appendix-lambdah}
To calculate the expected arrival rate of handoff calls,
we first need
to compute the probability $\alpha$ that a call terminates in a given
cell.   Let the probability density function (PDF) of the
dwell time $x_1$
and the PDF of the
call-holding time $x_2$
be
$f(x_{1}) ={h\mu} e^{ - h\mu x_{1}}$ and  $f(x_{2}) = \mu e^{ -\mu x_{1}
}$ respectively.
Since we assume
that $x_1$ and $x_2$ are independent, their joint PDF is given by
$f(x_{1},x_{2}) = h{\mu}^{2} e^{-( h\mu{x_{1}}+\mu x_{2})}$.
A call terminates in a cell if its dwell time in that cell is greater
than the remaining call duration, so a straightforward computation shows
that
\begin{eqnarray*}
\alpha & = & Prob(x_{1}>x_{2}) = \int_{0}^{\infty} (\int_{0}^{x_{1}} \mu
e^{- \mu x_{2}} dx_{2}) h\mu e^{-h\mu x_{1}}dx_{1}
= \frac{1}{h+1}
\end{eqnarray*}

We now compute the expected number of handoffs of a call.
There are two ways for a call to undergo exactly $j$ handoffs.   Either
there are $j$ successful handoffs and then the call is terminated or
there are $j-1$ successful handoffs and the last handoff is unsuccessful.
Thus, the probability that an accepted call
will undergo $j$ handoffs is
\commentout{

Let $P_{s}$, $ s = \bar{s_{0}},\cdots,\bar{s_{|S|-1}}$ be the steady
state probability of the Markov chain induced by the stationary policy
of the Markov Decision Process.

Define $P_{hd} = \sum_{s}^{}P_{s}P_{hd}^{s}$, $P_{cb} =
\sum_{s}^{}P_{s}P_{cb}^{s}$ where $P_{hd}^{s}$ is the handoff dropping
probability when in state $s$, $P_{cb}^{s}$ is the call blocking
probability when in state $s$.

The expected number of handoffs for a given call is
derived as follows:
\begin{eqnarray}
\ho &=& \sum_{s_0}^{}\sum_{s_{1}}^{} 1\times P({s_{t_0}=s_0})(1-\alpha)
P(s_{t_1}=s_1|s_{t_0}=s_0)[(1-P_{hd}^{s_{1}})\alpha+P_{hd}^{s_{1}}] +
\nonumber \\
& & \sum_{s_0}^{}\sum_{s_{1}}^{}\sum_{s_{2}}^{}
2\times P(s_{t_0}=s_0)(1-\alpha)P(s_{t_1}=s_1|s_{t_0}=s_0)(1-P_{hd}^{s_{1}})
P(s_{t_2}=s_{2}|s_{t_1}=s_1,s_{t_0}=s_0)
\nonumber \\
& & (1-\alpha)[(1-P_{hd}^{s_{2}})\alpha + P_{hd}^{s_{2}}] + \cdots
\nonumber \\
 &=& \sum_{s_0}^{} 1\times P({s_{t_0}=s_0})(1-\alpha)\sum_{s_{1}}^{}
P(s_{t_1}=s_1)[(1-P_{hd}^{s_{1}})\alpha+P_{hd}^{s_{1}}] +
\nonumber \\
& & \sum_{s_0}^{}
2\times P(s_{t_0}=s_0)(1-\alpha)\sum_{s_{1}}^{}P(s_{t_1}=s_1)(1-P_{hd}^{s_{1}})
\sum_{s_{2}}^{}P(s_{t_2}=s_{2})
(1-\alpha)[(1-P_{hd}^{s_{2}})\alpha + P_{hd}^{s_{2}}] + \cdots \nonumber \\
& = &  1 \times [(1-\alpha)(1-P_{hd})\alpha+(1-\alpha)P_{hd}] + \nonumber \\
& & 2\times(1-\alpha)^2 \sum_{s_{1}}^{} (1-P_{hd})^{s_{1}}P_{s_{1}}
\sum_{s_{2}}^{}((1-P_{hd}^{s_{2}})\alpha+ P_{hd}^{s_{2}})P_{s_{2}}
]+ \cdots
\nonumber \\
& =  &  1\times[(1-\alpha)(1-P_{hd})\alpha+ (1-\alpha)P_{hd}] +
2\times[(1-\alpha)^2 ((1-P_{hd})\alpha + P_{hd})(1-P_{hd})]
+ \cdots
\nonumber \\
& =  & \sum_{j}^{\infty}j[(1-\alpha)^{j}(1-P_{hd})^{j}\alpha+(1-\alpha)^{j}(1-P_{hd})^{j-1}P_{hd} ]
\nonumber \\
& = & \frac{(1-\alpha)}{\alpha+(1-\alpha)P_{hd}}
\nonumber \\
& = & \frac{h}{1+hP_{hd}}
\label{eq:h bar}
\end{eqnarray}
}

\( P_{j}=
(1-\alpha)^{j}(1-P_{hd})^{j}\alpha+(1-\alpha)^{j}(1-P_{hd})^{j-1}P_{hd}\),
where $P_{hd}$ is the handoff dropping probability of the system observed over time.
The expected number of handoffs
per call
is thus
\begin{equation}
\ho = \sum_{j=0}^{\infty}jP_{j}
   = \frac{(1-\alpha)}{\alpha+(1-\alpha)P_{hd}}
   = \frac{h}{1+hP_{hd}}
\label{eq:h bar}
\end{equation}

Because new calls are blocked with probability $P_{{cb}}$,
the arrival rate of new calls that are accepted is
$\lambda( 1-P_{cb} )$.
 Since the accepted new calls undergo
$\ho$ handoffs before it is either terminated or dropped,
the handoff arrival rate is as follows:
\begin{equation}
\lambda_{h} = \lambda\ho( 1-P_{cb} )
\label{eq:handoff rate}
\end{equation}

}

\section{Deriving the State-Transition Probabilities}
\label{appendix-transtionprob}

Given our experimental assumptions, 
the call holding times are exponentially distributed and the call
arrival times are determined by a Poisson distribution.
\cite{Guerin87} shows
experimentally
the handoff probability is also exponentially distributed;
thus, we make that assumption here, and calculate the parameters
experimentally, for each assumption $\vec{c}=(c_1,c_2)$ of the expected number of
call of each QoS class.

 By taking the minimum of all these distributions,
we can compute
the PDF describing the time that the next event happens in state
$x=(x_1,x_2,\calevts_{x})$.
This PDF is defined as
\begin{equation}
f(t) = \omega e^{-\omega t},
\end{equation}
where
$\omega = (x_{1}+x_{2})\mu+\lambda_1+\lambda_2+(c_1+c_2)\rho\mu$,
$\lambda_1 = R_{dv}\lambda$, $\lambda_2 = (1-R_{dv})\lambda$;

So the expected time until a new event occurs when in state $x$ is
\begin{eqnarray}
\label{eq:tau}
\tau(x) &=& \int_{0}^{\infty}t f(t) d t
\nonumber \\
 &=& \left[ \sum_{i=1}^{2}(x_{i}\mu_{i}+\lambda_{i} + c_i\rho\mu)
\right]^{-1}
\end{eqnarray}

The probability
of a transition from state $x$ to state $y$ given action
$a$ is:
\begin{equation}
P_{xy}^{a}= \left\{ \begin{array}{ll}
  \lambda_1\tau(x) & \mbox{$y = (x_1+\sigma_1,x_2+\sigma_2,(\arevt,1))$} \\
  \lambda_2\tau(x) & \mbox{$y = (x_1+\sigma_1,x_2+\sigma_2,(\arevt,2))$} \\
  c_1\rho\mu\tau(x)& \mbox{$y = (x_1+\sigma_1,x_2+\sigma_2,(h,1))$} \\
  c_2\rho\mu\tau(x)& \mbox{$y = (x_1+\sigma_1,x_2+\sigma_2,(h,2))$} \\
  x_{1}\mu\tau(x)  & \mbox{$y = (x_1+\sigma_1,x_2+\sigma_2,(d,1))$} \\
  x_{2}\mu\tau(x)  & \mbox{$y = (x_1+\sigma_1,x_2+\sigma_2,(d,2))$} \\
  \end{array}
  \right.
\end{equation}
where
\begin{equation}
\sigma_i = \left\{ \begin{array}{ll}
 1 & \mbox{if  $a=accept$ and $\theta_x = (\arevt,i)$ or $(h,i)$} \\
 -1& \mbox{if $\theta_x = (d,i)$} \\
 0 & \mbox{otherwise.}
\end{array}
 \right.
\end{equation}

\commentout{
\section{Deriving $P_{h}$ and $P_{s}$ for NAG}
\label{appendix-phps}
Let $P_{e}$ denote the probability that a
call terminates
time $T_{est}$.  The probability $P_{h}$ that a
call is handed off
to a given neighboring cell in the next $T_{est}$ time units
is one sixth of the probability that $T_{est}$ is greater than the
call's dwell time, and can be calculated as follows:
\begin{eqnarray}
P_{h} &=& Prob(t<T_{est})/6
\nonumber \\
&=& \frac{1}{6} \int_{0}^{T_{est}} h\mu e^{-h\mu t} d t
\nonumber \\
& = & (1- e^{-h\mu T_{est}})/6
\end{eqnarray}

The probability $P_e$ that a
call terminates in the current cell is the
probability that $T_{est}$ is greater than the remaining call holding
time:
\begin{eqnarray}
P_{e} &=& Prob(t<T_{est})
\nonumber \\
&=& \int_{0}^{T_{est}} \mu e^{-\mu t} d t
\nonumber \\
&= & 1- e^{-\mu T_{est}}
\end{eqnarray}

The probability $P_{s}$ that a mobile will stay in the current cell
during the next $T_{est}$ time units is the probability that the
call is not handed off and does not terminate in time $T_{est}$.

\begin{eqnarray}
P_{h} &=& (1-P_e)(1-6 \times P_{h})
\nonumber \\
&= & e^{-(h+1)\mu T_{est}}
\end{eqnarray}
}

\commentout{
\section{The Convergence of the Modified Value Iteration Algorithm}
\label{appendix-converge}

We give a sketch of the proof.
The blocking and dropping probabilities
$P_{hd}^{m+1}$ and $P_{cb}^{m+1}$ can be expressed as some function $g$
of the $P_{hd}^m$ and $P_{cb}^m$.
The function $g$ can be computed using value iteration.
The function $g$ intersects the function \ref{eq:handoff rate}.
Standard algorithms for finding one of the intersections of two
functions that converges can be applied.

}

\end{document}